\documentclass[a4paper,10pt,journal]{IEEEtran}
\usepackage{cite}
\usepackage{siunitx}
\DeclareSIUnit \GHz {GHz}
\usepackage{graphicx}
\usepackage{amsfonts}
\usepackage{amsmath}
\usepackage[caption=false,font=footnotesize]{subfig}
\usepackage[hyphens]{url}
\usepackage{algorithm}
\usepackage{algpseudocode}

\newcommand{\FF}[1]{{\mathbb{F}}}

\usepackage{tikz}
\usetikzlibrary{arrows,automata}

\usepackage{enumerate}

\newtheorem{definition}{Definition}[section]
\newtheorem{lemma}{Lemma}[section]
\newtheorem{theorem}{Theorem}[section]

\newtheorem{remark}{Remark}[section]
\newtheorem{example}{Example}[section]
\newcolumntype{P}[1]{>{\centering\arraybackslash}p{#1}}
\newcolumntype{M}[1]{>{\centering\arraybackslash}m{#1}}
\begin{document}
\title{{On Intercept Probability Minimization under\\ Sparse Random Linear Network Coding}}

\author{Andrea Tassi, Robert J. Piechocki, and Andrew Nix
\thanks{
Copyright (c) 2015 IEEE. Personal use of this material is permitted. However, permission to use this material for any other purposes must be obtained from the IEEE by sending a request to {\tt pubs-permissions@ieee.org}.

A. Tassi, R. J. Piechocki and A. Nix are with the Department of Electrical and Electronic Engineering, University of Bristol, UK (e-mail: {\tt \{A.Tassi,R.J.Piechocki,Andy.Nix\}@bristol.ac.uk}).
} \vspace*{-4mm}}

\maketitle

\begin{abstract}
This paper considers a network where a node wishes to transmit a source message to a legitimate receiver in the presence of an eavesdropper. The transmitter secures its transmissions employing a sparse implementation of Random Linear Network Coding (RLNC). A tight approximation to the probability of the eavesdropper recovering the source message is provided. The proposed approximation applies to both the cases where transmissions occur without feedback or where the reliability of the feedback channel is impaired by an eavesdropper jamming the feedback channel.
An optimization framework for minimizing the intercept probability by optimizing the sparsity of the RLNC is also presented. Results validate the proposed approximation and quantify the gain provided by our optimization over solutions where non-sparse RLNC is used.
\end{abstract}


\vspace{-5mm}\section{Introduction}{
Due to the broadcast nature of the medium, wireless communications can be vulnerable to eavesdropping. Physical layer security strategies, operating at the lower protocol stack layers, aim to achieve the secrecy of transmitted messages. In partibular, an eavesdropper is prevented from recovering any of the packets broadcast by a source node (\emph{per-packet secrecy}) by optimizing the transmission rate~\cite{4529264}.

In this paper, we advance and compare against the framework for physical layer security presented in~\cite{6777406}, and more recently in~\cite{7214217}. In particular, we refer to a system model where achieving per-packet secrecy is not necessary if the transmitted packets are a function of a source message intended to be delivered to a legitimate receiver, and if, in order to recover the source message, a receiver has to collect at least a target number of packets~\cite{1023595}. As observed in~\cite{6777406} and~\cite{7214217}, this assumption is met by Random Linear Network Coding (RLNC)~\cite{8281108}, where a source node generates a stream of coded packets by linearly combining the source packets forming a source message. The legitimate receiver or an eavesdropper can recover the source message only if they successfully receive a number of linearly independent coded packets equal to the number of source packets defining the source message.

We secure communications by minimizing the intercept probability -- defined as the probability of an eavesdropper recovering the source message intended for a legitimate receiver. Unlike~\cite{6777406,7214217}, the devised proposal applies to both the case when the legitimate receiver does and does not acknowledge the source the successful reception of a message. This is achieved, by establishing our theoretical framework under the conditions where the transmission of acknowledgment messages takes place over a feedback channel that is not assumed fully reliable.
In particular, our performance investigation will focus on attacks where an eavesdropper attempts to increase its intercept probability by jamming the feedback channel -- thus, increasing the probability of the acknowledgment message not being successfully received and forcing the source node to keep transmitting coded packets even after the legitimate receiver successfully recovered a source message.
To avoid that, we will show how the intercept probability can be significantly reduced by adopting a sparse implementation of the RLNC approach where the number of non-zero elements in the encoding matrix is smaller than in the case of classic RLNC~\cite{7335581}.

In this paper, we provide the following key contributions:
\begin{itemize}
\item 
Existing expressions of the intercept probability are only applicable to extreme cases where the legitimate receiver either does not acknowledge to the source the successful reception of a source message or when an acknowledgment message is transmitted over a fully reliable feedback channel. By resorting to a novel Markov chain-based model, we propose a generic approximation of the intercept probability that is also applicable when the feedback channel is impaired by an arbitrary erasure probability.
\item  By employing a sparse implementation of RLNC, we devise a novel optimization strategy for optimizing the sparsity of the code and then minimizing the intercept probability when the feedback channel is jammed.
\end{itemize}}

The rest of the paper is organized as follows. Section~\ref{sec.SM} describes the considered system model. Section~\ref{sec.PA} presents our novel approximation of the intercept probability and Section~\ref{sec.OM} shows how the sparsity of the code can be optimized to minimize the intercept probability. The accuracy of the proposed approximation and the effectiveness of our optimization model are presented in Section~\ref{sec.NR}. Finally, in Section~\ref{sec.CL}, we draw our conclusions.

\vspace{-2mm}\section{System Model}\label{sec.SM}
We consider a system model where a node (Alice) wishes to transmit to a legitimate receiving node (Bob) a source message in the presence of an eavesdropper (Eve), over a broadcast channel. Bob and Eve experience a packet error probability equal to $\epsilon_{\mathrm{B}}$ and $\epsilon_{\mathrm{E}}$, respectively.

{We assume that the packet erasures experienced by Bob and Eve occur as statistically independent events and, based on a general condition for physical layer security over a Wyner's wiretap channel model~\cite[Chapter~1]{bloch_barros_2011}, $\epsilon_{\mathrm{B}} \leq \epsilon_{\mathrm{E}}$~\cite{1055917}.
\begin{remark}
It directly follows from~\cite{6777406,7214217} that, for $\epsilon_{\mathrm{B}} > \epsilon_{\mathrm{E}}$, the average number of coded packet successfully received by Bob is smaller than that received by Eve -- thus, the average number of coded packet transmissions that Eve needs to recover a source message is inevitably smaller than the number of coded packets Bob needs to recover a source message. That is, for $\epsilon_{\mathrm{B}} > \epsilon_{\mathrm{E}}$, the secrecy capacity of a multicast or broadcast communication system cannot be improved by only employing strategies based on rateless codes. Thus, alternative physical layer security techniques achieving per-packet secrecy have to be used. The investigation of scenarios where $\epsilon_{\mathrm{B}} > \epsilon_{\mathrm{E}}$ are beyond the scope of this paper.
\end{remark}}

Alice segments the source message into $K$ source packets and linearly combines at random the source packets to obtain $\Hat{N}$ coded packets for transmission according to the sparse RLNC principle defined as follows.
\begin{definition}{
Each coded packet $\mathbf{c}_j$ is obtained as $\mathbf{c}_j = \sum_{i = 1}^K g_{i,j} \cdot \mathbf{s}_i$, where $g_{i,j}$ follows the following probability law~\cite{7335581}:
\vspace{-3mm}\begin{equation}\label{eq.pl}
  \mathbb{P}\left(g_{i,j} = v\right) = \left\{ 
  \begin{array}{l l}
      p & \quad \text{if $v = 0$}\\
    \displaystyle\frac{1-p}{q-1} & \quad \text{otherwise,}\\
  \end{array} \right.
\end{equation}
where $\frac{1}{q} < p < 1$ and $q$ is the size of the finite field $\mathbb{F}_q$ over which network coding operations are performed. The bigger $p$, the more likely that $g_{i,j}$ is equal to $0$. Thus, the average number of source packets concurring in the generation of a coded packet is a function of $p$. Classic RLNC assumes \mbox{$p = \frac{1}{q}$}~\cite{8248799}.}
\end{definition}

Let $n_\mathrm{B}$ and $n_\mathrm{E}$ be the number of  coded packets successfully received by Bob and Eve, for $0 \leq n_\mathrm{B} \leq \Hat{N}$ and $0 \leq n_\mathrm{E} \leq \Hat{N}$, respectively.
Column by column, Bob and Eve populate a $K \times n_{\mathrm{B}}$ and a $K \times n_{\mathrm{E}}$ decoding matrix $\mathbf{M}_{\mathrm{B}}$ and $\mathbf{M}_{\mathrm{E}}$, respectively, with the coding vectors associated with the coded packets they successfully received.
Bob and Eve recover the source message as soon as the defect of the decoding matrix, defined as $\mathrm{def}(\mathbf{M}_{\mathrm{X}}) = K - \mathrm{rank}(\mathbf{M}_{\mathrm{X}})$ is equal to zero, for $X = \mathrm{B}$ and $X = \mathrm{E}$, respectively~\cite{8281108}.

As soon as the source message has been successfully recovered, Bob transmits an acknowledgment message to Alice over a feedback channel. Alice stops broadcasting coded packets as soon as the feedback is successfully received or when $\Hat{N} > K$ coded packets have been broadcast.
The acknowledgment message is re-transmitted when Bob detects a new coded packet transmission pertaining to a source message that Bob has already recovered. The detection of new packet transmissions is assumed to be fully reliable.
The feedback channel is assumed independent and separated from the broadcast channel used to transmit coded packets. The erasures of acknowledgement messages occur with probability $\epsilon_{\mathrm{K}}$, for $0 < \epsilon_{\mathrm{K}} \leq 1$.

\vspace{-5mm}\section{Performance Analysis}\label{sec.PA}
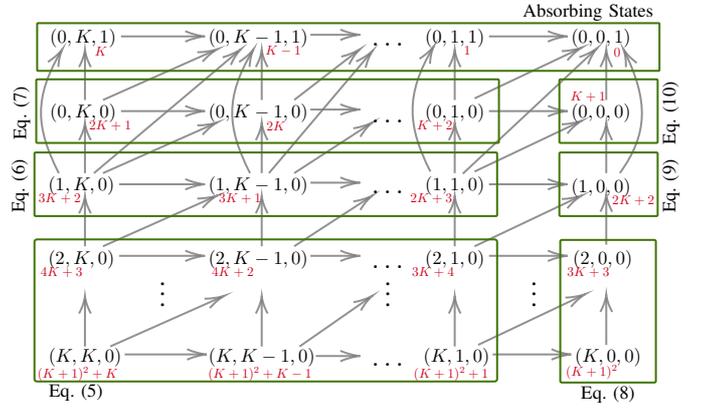
\begin{figure}[tb]
\vspace{-8mm}
    \begin{center}
\tikzset{every picture/.style={line width=0.75pt}} 

\begin{tikzpicture}[x=0.75pt,y=0.75pt,yscale=0.8,xscale=0.9]

\draw [color={rgb, 255:red, 126; green, 126; blue, 126 }  ,draw opacity=0.85 ]   (190,141) -- (235.52,141) ;
\draw [shift={(237.52,141)}, rotate = 180] [color={rgb, 255:red, 126; green, 126; blue, 126 }  ,draw opacity=0.85 ][line width=0.75]    (10.93,-3.29) .. controls (6.95,-1.4) and (3.31,-0.3) .. (0,0) .. controls (3.31,0.3) and (6.95,1.4) .. (10.93,3.29)   ;

\draw [color={rgb, 255:red, 126; green, 126; blue, 126 }  ,draw opacity=0.85 ]   (190,186) -- (235.52,186) ;
\draw [shift={(237.52,186)}, rotate = 180] [color={rgb, 255:red, 126; green, 126; blue, 126 }  ,draw opacity=0.85 ][line width=0.75]    (10.93,-3.29) .. controls (6.95,-1.4) and (3.31,-0.3) .. (0,0) .. controls (3.31,0.3) and (6.95,1.4) .. (10.93,3.29)   ;

\draw [color={rgb, 255:red, 126; green, 126; blue, 126 }  ,draw opacity=0.85 ]   (191,232) -- (235.52,232) ;
\draw [shift={(237.52,232)}, rotate = 180] [color={rgb, 255:red, 126; green, 126; blue, 126 }  ,draw opacity=0.85 ][line width=0.75]    (10.93,-3.29) .. controls (6.95,-1.4) and (3.31,-0.3) .. (0,0) .. controls (3.31,0.3) and (6.95,1.4) .. (10.93,3.29)   ;

\draw [color={rgb, 255:red, 126; green, 126; blue, 126 }  ,draw opacity=0.85 ]   (191,279) -- (235.52,279) ;
\draw [shift={(237.52,279)}, rotate = 180] [color={rgb, 255:red, 126; green, 126; blue, 126 }  ,draw opacity=0.85 ][line width=0.75]    (10.93,-3.29) .. controls (6.95,-1.4) and (3.31,-0.3) .. (0,0) .. controls (3.31,0.3) and (6.95,1.4) .. (10.93,3.29)   ;

\draw [color={rgb, 255:red, 126; green, 126; blue, 126 }  ,draw opacity=0.85 ]   (193,79) -- (236.52,79) ;
\draw [shift={(238.52,79)}, rotate = 180] [color={rgb, 255:red, 126; green, 126; blue, 126 }  ,draw opacity=0.85 ][line width=0.75]    (10.93,-3.29) .. controls (6.95,-1.4) and (3.31,-0.3) .. (0,0) .. controls (3.31,0.3) and (6.95,1.4) .. (10.93,3.29)   ;

\draw [color={rgb, 255:red, 126; green, 126; blue, 126 }  ,draw opacity=0.85 ]   (299,78) -- (321.52,78) ;
\draw [shift={(323.52,78)}, rotate = 180] [color={rgb, 255:red, 126; green, 126; blue, 126 }  ,draw opacity=0.85 ][line width=0.75]    (10.93,-3.29) .. controls (6.95,-1.4) and (3.31,-0.3) .. (0,0) .. controls (3.31,0.3) and (6.95,1.4) .. (10.93,3.29)   ;

\draw [color={rgb, 255:red, 126; green, 126; blue, 126 }  ,draw opacity=0.85 ]   (296,141) -- (321.52,141) ;
\draw [shift={(323.52,141)}, rotate = 180] [color={rgb, 255:red, 126; green, 126; blue, 126 }  ,draw opacity=0.85 ][line width=0.75]    (10.93,-3.29) .. controls (6.95,-1.4) and (3.31,-0.3) .. (0,0) .. controls (3.31,0.3) and (6.95,1.4) .. (10.93,3.29)   ;

\draw [color={rgb, 255:red, 126; green, 126; blue, 126 }  ,draw opacity=0.85 ]   (296,187) -- (321.52,187) ;
\draw [shift={(323.52,187)}, rotate = 180] [color={rgb, 255:red, 126; green, 126; blue, 126 }  ,draw opacity=0.85 ][line width=0.75]    (10.93,-3.29) .. controls (6.95,-1.4) and (3.31,-0.3) .. (0,0) .. controls (3.31,0.3) and (6.95,1.4) .. (10.93,3.29)   ;

\draw [color={rgb, 255:red, 126; green, 126; blue, 126 }  ,draw opacity=0.85 ]   (296,233) -- (321.52,233) ;
\draw [shift={(323.52,233)}, rotate = 180] [color={rgb, 255:red, 126; green, 126; blue, 126 }  ,draw opacity=0.85 ][line width=0.75]    (10.93,-3.29) .. controls (6.95,-1.4) and (3.31,-0.3) .. (0,0) .. controls (3.31,0.3) and (6.95,1.4) .. (10.93,3.29)   ;

\draw [color={rgb, 255:red, 126; green, 126; blue, 126 }  ,draw opacity=0.85 ]   (296,280) -- (321.52,280) ;
\draw [shift={(323.52,280)}, rotate = 180] [color={rgb, 255:red, 126; green, 126; blue, 126 }  ,draw opacity=0.85 ][line width=0.75]    (10.93,-3.29) .. controls (6.95,-1.4) and (3.31,-0.3) .. (0,0) .. controls (3.31,0.3) and (6.95,1.4) .. (10.93,3.29)   ;

\draw [color={rgb, 255:red, 126; green, 126; blue, 126 }  ,draw opacity=0.85 ]   (395,141) -- (436.52,141) ;
\draw [shift={(438.52,141)}, rotate = 180] [color={rgb, 255:red, 126; green, 126; blue, 126 }  ,draw opacity=0.85 ][line width=0.75]    (10.93,-3.29) .. controls (6.95,-1.4) and (3.31,-0.3) .. (0,0) .. controls (3.31,0.3) and (6.95,1.4) .. (10.93,3.29)   ;

\draw [color={rgb, 255:red, 126; green, 126; blue, 126 }  ,draw opacity=0.85 ]   (395,186) -- (436.52,186) ;
\draw [shift={(438.52,186)}, rotate = 180] [color={rgb, 255:red, 126; green, 126; blue, 126 }  ,draw opacity=0.85 ][line width=0.75]    (10.93,-3.29) .. controls (6.95,-1.4) and (3.31,-0.3) .. (0,0) .. controls (3.31,0.3) and (6.95,1.4) .. (10.93,3.29)   ;

\draw [color={rgb, 255:red, 126; green, 126; blue, 126 }  ,draw opacity=0.85 ]   (395,232) -- (436.52,232) ;
\draw [shift={(438.52,232)}, rotate = 180] [color={rgb, 255:red, 126; green, 126; blue, 126 }  ,draw opacity=0.85 ][line width=0.75]    (10.93,-3.29) .. controls (6.95,-1.4) and (3.31,-0.3) .. (0,0) .. controls (3.31,0.3) and (6.95,1.4) .. (10.93,3.29)   ;

\draw [color={rgb, 255:red, 126; green, 126; blue, 126 }  ,draw opacity=0.85 ]   (395,279) -- (436.52,279) ;
\draw [shift={(438.52,279)}, rotate = 180] [color={rgb, 255:red, 126; green, 126; blue, 126 }  ,draw opacity=0.85 ][line width=0.75]    (10.93,-3.29) .. controls (6.95,-1.4) and (3.31,-0.3) .. (0,0) .. controls (3.31,0.3) and (6.95,1.4) .. (10.93,3.29)   ;

\draw [color={rgb, 255:red, 126; green, 126; blue, 126 }  ,draw opacity=0.85 ]   (397,79) -- (438.52,79) ;
\draw [shift={(440.52,79)}, rotate = 180] [color={rgb, 255:red, 126; green, 126; blue, 126 }  ,draw opacity=0.85 ][line width=0.75]    (10.93,-3.29) .. controls (6.95,-1.4) and (3.31,-0.3) .. (0,0) .. controls (3.31,0.3) and (6.95,1.4) .. (10.93,3.29)   ;

\draw [color={rgb, 255:red, 126; green, 126; blue, 126 }  ,draw opacity=0.85 ]   (171,84) -- (171,113.67) ;
\draw [shift={(171,115.67)}, rotate = 270] [color={rgb, 255:red, 126; green, 126; blue, 126 }  ,draw opacity=0.85 ][line width=0.75]    (10.93,-3.29) .. controls (6.95,-1.4) and (3.31,-0.3) .. (0,0) .. controls (3.31,0.3) and (6.95,1.4) .. (10.93,3.29)   ;

\draw [color={rgb, 255:red, 126; green, 126; blue, 126 }  ,draw opacity=0.85 ]   (269,84) -- (269,113.67) ;
\draw [shift={(269,115.67)}, rotate = 270] [color={rgb, 255:red, 126; green, 126; blue, 126 }  ,draw opacity=0.85 ][line width=0.75]    (10.93,-3.29) .. controls (6.95,-1.4) and (3.31,-0.3) .. (0,0) .. controls (3.31,0.3) and (6.95,1.4) .. (10.93,3.29)   ;

\draw [color={rgb, 255:red, 126; green, 126; blue, 126 }  ,draw opacity=0.85 ]   (376,84) -- (376,113.67) ;
\draw [shift={(376,115.67)}, rotate = 270] [color={rgb, 255:red, 126; green, 126; blue, 126 }  ,draw opacity=0.85 ][line width=0.75]    (10.93,-3.29) .. controls (6.95,-1.4) and (3.31,-0.3) .. (0,0) .. controls (3.31,0.3) and (6.95,1.4) .. (10.93,3.29)   ;

\draw [color={rgb, 255:red, 126; green, 126; blue, 126 }  ,draw opacity=0.85 ]   (460,85) -- (460,114.67) ;
\draw [shift={(460,116.67)}, rotate = 270] [color={rgb, 255:red, 126; green, 126; blue, 126 }  ,draw opacity=0.85 ][line width=0.75]    (10.93,-3.29) .. controls (6.95,-1.4) and (3.31,-0.3) .. (0,0) .. controls (3.31,0.3) and (6.95,1.4) .. (10.93,3.29)   ;

\draw [color={rgb, 255:red, 126; green, 126; blue, 126 }  ,draw opacity=0.85 ]   (185,84) -- (246.71,115.1) ;
\draw [shift={(248.5,116)}, rotate = 206.75] [color={rgb, 255:red, 126; green, 126; blue, 126 }  ,draw opacity=0.85 ][line width=0.75]    (10.93,-3.29) .. controls (6.95,-1.4) and (3.31,-0.3) .. (0,0) .. controls (3.31,0.3) and (6.95,1.4) .. (10.93,3.29)   ;

\draw [color={rgb, 255:red, 126; green, 126; blue, 126 }  ,draw opacity=0.85 ]   (290,84) -- (328.93,114.76) ;
\draw [shift={(330.5,116)}, rotate = 218.31] [color={rgb, 255:red, 126; green, 126; blue, 126 }  ,draw opacity=0.85 ][line width=0.75]    (10.93,-3.29) .. controls (6.95,-1.4) and (3.31,-0.3) .. (0,0) .. controls (3.31,0.3) and (6.95,1.4) .. (10.93,3.29)   ;

\draw [color={rgb, 255:red, 126; green, 126; blue, 126 }  ,draw opacity=0.85 ]   (389,85) -- (446.74,116.05) ;
\draw [shift={(448.5,117)}, rotate = 208.27] [color={rgb, 255:red, 126; green, 126; blue, 126 }  ,draw opacity=0.85 ][line width=0.75]    (10.93,-3.29) .. controls (6.95,-1.4) and (3.31,-0.3) .. (0,0) .. controls (3.31,0.3) and (6.95,1.4) .. (10.93,3.29)   ;

\draw [color={rgb, 255:red, 126; green, 126; blue, 126 }  ,draw opacity=0.85 ]   (171,147) -- (171,176.67) ;
\draw [shift={(171,178.67)}, rotate = 270] [color={rgb, 255:red, 126; green, 126; blue, 126 }  ,draw opacity=0.85 ][line width=0.75]    (10.93,-3.29) .. controls (6.95,-1.4) and (3.31,-0.3) .. (0,0) .. controls (3.31,0.3) and (6.95,1.4) .. (10.93,3.29)   ;

\draw [color={rgb, 255:red, 126; green, 126; blue, 126 }  ,draw opacity=0.85 ]   (269,147) -- (269,176.67) ;
\draw [shift={(269,178.67)}, rotate = 270] [color={rgb, 255:red, 126; green, 126; blue, 126 }  ,draw opacity=0.85 ][line width=0.75]    (10.93,-3.29) .. controls (6.95,-1.4) and (3.31,-0.3) .. (0,0) .. controls (3.31,0.3) and (6.95,1.4) .. (10.93,3.29)   ;

\draw [color={rgb, 255:red, 126; green, 126; blue, 126 }  ,draw opacity=0.85 ]   (376,147) -- (376,176.67) ;
\draw [shift={(376,178.67)}, rotate = 270] [color={rgb, 255:red, 126; green, 126; blue, 126 }  ,draw opacity=0.85 ][line width=0.75]    (10.93,-3.29) .. controls (6.95,-1.4) and (3.31,-0.3) .. (0,0) .. controls (3.31,0.3) and (6.95,1.4) .. (10.93,3.29)   ;

\draw [color={rgb, 255:red, 126; green, 126; blue, 126 }  ,draw opacity=0.85 ]   (460,148) -- (460,177.67) ;
\draw [shift={(460,179.67)}, rotate = 270] [color={rgb, 255:red, 126; green, 126; blue, 126 }  ,draw opacity=0.85 ][line width=0.75]    (10.93,-3.29) .. controls (6.95,-1.4) and (3.31,-0.3) .. (0,0) .. controls (3.31,0.3) and (6.95,1.4) .. (10.93,3.29)   ;

\draw [color={rgb, 255:red, 126; green, 126; blue, 126 }  ,draw opacity=0.85 ]   (181,147) -- (242.71,178.1) ;
\draw [shift={(244.5,179)}, rotate = 206.75] [color={rgb, 255:red, 126; green, 126; blue, 126 }  ,draw opacity=0.85 ][line width=0.75]    (10.93,-3.29) .. controls (6.95,-1.4) and (3.31,-0.3) .. (0,0) .. controls (3.31,0.3) and (6.95,1.4) .. (10.93,3.29)   ;

\draw [color={rgb, 255:red, 126; green, 126; blue, 126 }  ,draw opacity=0.85 ]   (287,147) -- (328.89,177.81) ;
\draw [shift={(330.5,179)}, rotate = 216.34] [color={rgb, 255:red, 126; green, 126; blue, 126 }  ,draw opacity=0.85 ][line width=0.75]    (10.93,-3.29) .. controls (6.95,-1.4) and (3.31,-0.3) .. (0,0) .. controls (3.31,0.3) and (6.95,1.4) .. (10.93,3.29)   ;

\draw [color={rgb, 255:red, 126; green, 126; blue, 126 }  ,draw opacity=0.85 ]   (387,147) -- (444.74,178.05) ;
\draw [shift={(446.5,179)}, rotate = 208.27] [color={rgb, 255:red, 126; green, 126; blue, 126 }  ,draw opacity=0.85 ][line width=0.75]    (10.93,-3.29) .. controls (6.95,-1.4) and (3.31,-0.3) .. (0,0) .. controls (3.31,0.3) and (6.95,1.4) .. (10.93,3.29)   ;

\draw [color={rgb, 255:red, 126; green, 126; blue, 126 }  ,draw opacity=0.85 ]   (171,193) -- (171,222.67) ;
\draw [shift={(171,224.67)}, rotate = 270] [color={rgb, 255:red, 126; green, 126; blue, 126 }  ,draw opacity=0.85 ][line width=0.75]    (10.93,-3.29) .. controls (6.95,-1.4) and (3.31,-0.3) .. (0,0) .. controls (3.31,0.3) and (6.95,1.4) .. (10.93,3.29)   ;

\draw [color={rgb, 255:red, 126; green, 126; blue, 126 }  ,draw opacity=0.85 ]   (269,193) -- (269,222.67) ;
\draw [shift={(269,224.67)}, rotate = 270] [color={rgb, 255:red, 126; green, 126; blue, 126 }  ,draw opacity=0.85 ][line width=0.75]    (10.93,-3.29) .. controls (6.95,-1.4) and (3.31,-0.3) .. (0,0) .. controls (3.31,0.3) and (6.95,1.4) .. (10.93,3.29)   ;

\draw [color={rgb, 255:red, 126; green, 126; blue, 126 }  ,draw opacity=0.85 ]   (376,193) -- (376,222.67) ;
\draw [shift={(376,224.67)}, rotate = 270] [color={rgb, 255:red, 126; green, 126; blue, 126 }  ,draw opacity=0.85 ][line width=0.75]    (10.93,-3.29) .. controls (6.95,-1.4) and (3.31,-0.3) .. (0,0) .. controls (3.31,0.3) and (6.95,1.4) .. (10.93,3.29)   ;

\draw [color={rgb, 255:red, 126; green, 126; blue, 126 }  ,draw opacity=0.85 ]   (460,193) -- (460,222.67) ;
\draw [shift={(460,224.67)}, rotate = 270] [color={rgb, 255:red, 126; green, 126; blue, 126 }  ,draw opacity=0.85 ][line width=0.75]    (10.93,-3.29) .. controls (6.95,-1.4) and (3.31,-0.3) .. (0,0) .. controls (3.31,0.3) and (6.95,1.4) .. (10.93,3.29)   ;

\draw [color={rgb, 255:red, 126; green, 126; blue, 126 }  ,draw opacity=0.85 ]   (181,193) -- (242.71,224.1) ;
\draw [shift={(244.5,225)}, rotate = 206.75] [color={rgb, 255:red, 126; green, 126; blue, 126 }  ,draw opacity=0.85 ][line width=0.75]    (10.93,-3.29) .. controls (6.95,-1.4) and (3.31,-0.3) .. (0,0) .. controls (3.31,0.3) and (6.95,1.4) .. (10.93,3.29)   ;

\draw [color={rgb, 255:red, 126; green, 126; blue, 126 }  ,draw opacity=0.85 ]   (287,193) -- (328.89,223.81) ;
\draw [shift={(330.5,225)}, rotate = 216.34] [color={rgb, 255:red, 126; green, 126; blue, 126 }  ,draw opacity=0.85 ][line width=0.75]    (10.93,-3.29) .. controls (6.95,-1.4) and (3.31,-0.3) .. (0,0) .. controls (3.31,0.3) and (6.95,1.4) .. (10.93,3.29)   ;

\draw [color={rgb, 255:red, 126; green, 126; blue, 126 }  ,draw opacity=0.85 ]   (387,194) -- (444.74,225.05) ;
\draw [shift={(446.5,226)}, rotate = 208.27] [color={rgb, 255:red, 126; green, 126; blue, 126 }  ,draw opacity=0.85 ][line width=0.75]    (10.93,-3.29) .. controls (6.95,-1.4) and (3.31,-0.3) .. (0,0) .. controls (3.31,0.3) and (6.95,1.4) .. (10.93,3.29)   ;

\draw [color={rgb, 255:red, 126; green, 126; blue, 126 }  ,draw opacity=0.85 ]   (171,239) -- (171,268.67) ;
\draw [shift={(171,270.67)}, rotate = 270] [color={rgb, 255:red, 126; green, 126; blue, 126 }  ,draw opacity=0.85 ][line width=0.75]    (10.93,-3.29) .. controls (6.95,-1.4) and (3.31,-0.3) .. (0,0) .. controls (3.31,0.3) and (6.95,1.4) .. (10.93,3.29)   ;

\draw [color={rgb, 255:red, 126; green, 126; blue, 126 }  ,draw opacity=0.85 ]   (269,239) -- (269,268.67) ;
\draw [shift={(269,270.67)}, rotate = 270] [color={rgb, 255:red, 126; green, 126; blue, 126 }  ,draw opacity=0.85 ][line width=0.75]    (10.93,-3.29) .. controls (6.95,-1.4) and (3.31,-0.3) .. (0,0) .. controls (3.31,0.3) and (6.95,1.4) .. (10.93,3.29)   ;

\draw [color={rgb, 255:red, 126; green, 126; blue, 126 }  ,draw opacity=0.85 ]   (376,239) -- (376,268.67) ;
\draw [shift={(376,270.67)}, rotate = 270] [color={rgb, 255:red, 126; green, 126; blue, 126 }  ,draw opacity=0.85 ][line width=0.75]    (10.93,-3.29) .. controls (6.95,-1.4) and (3.31,-0.3) .. (0,0) .. controls (3.31,0.3) and (6.95,1.4) .. (10.93,3.29)   ;

\draw [color={rgb, 255:red, 126; green, 126; blue, 126 }  ,draw opacity=0.85 ]   (460,239) -- (460,268.67) ;
\draw [shift={(460,270.67)}, rotate = 270] [color={rgb, 255:red, 126; green, 126; blue, 126 }  ,draw opacity=0.85 ][line width=0.75]    (10.93,-3.29) .. controls (6.95,-1.4) and (3.31,-0.3) .. (0,0) .. controls (3.31,0.3) and (6.95,1.4) .. (10.93,3.29)   ;

\draw [color={rgb, 255:red, 126; green, 126; blue, 126 }  ,draw opacity=0.85 ]   (181,239) -- (242.71,270.1) ;
\draw [shift={(244.5,271)}, rotate = 206.75] [color={rgb, 255:red, 126; green, 126; blue, 126 }  ,draw opacity=0.85 ][line width=0.75]    (10.93,-3.29) .. controls (6.95,-1.4) and (3.31,-0.3) .. (0,0) .. controls (3.31,0.3) and (6.95,1.4) .. (10.93,3.29)   ;

\draw [color={rgb, 255:red, 126; green, 126; blue, 126 }  ,draw opacity=0.85 ]   (287,239) -- (322,269.68) ;
\draw [shift={(323.5,271)}, rotate = 221.24] [color={rgb, 255:red, 126; green, 126; blue, 126 }  ,draw opacity=0.85 ][line width=0.75]    (10.93,-3.29) .. controls (6.95,-1.4) and (3.31,-0.3) .. (0,0) .. controls (3.31,0.3) and (6.95,1.4) .. (10.93,3.29)   ;

\draw [color={rgb, 255:red, 126; green, 126; blue, 126 }  ,draw opacity=0.85 ]   (387,240) -- (444.74,271.05) ;
\draw [shift={(446.5,272)}, rotate = 208.27] [color={rgb, 255:red, 126; green, 126; blue, 126 }  ,draw opacity=0.85 ][line width=0.75]    (10.93,-3.29) .. controls (6.95,-1.4) and (3.31,-0.3) .. (0,0) .. controls (3.31,0.3) and (6.95,1.4) .. (10.93,3.29)   ;

\draw [color={rgb, 255:red, 126; green, 126; blue, 126 }  ,draw opacity=0.85 ]   (176,193) -- (256.06,269.62) ;
\draw [shift={(257.5,271)}, rotate = 223.74] [color={rgb, 255:red, 126; green, 126; blue, 126 }  ,draw opacity=0.85 ][line width=0.75]    (10.93,-3.29) .. controls (6.95,-1.4) and (3.31,-0.3) .. (0,0) .. controls (3.31,0.3) and (6.95,1.4) .. (10.93,3.29)   ;

\draw [color={rgb, 255:red, 126; green, 126; blue, 126 }  ,draw opacity=0.85 ]   (273,193) -- (329.3,268.4) ;
\draw [shift={(330.5,270)}, rotate = 233.25] [color={rgb, 255:red, 126; green, 126; blue, 126 }  ,draw opacity=0.85 ][line width=0.75]    (10.93,-3.29) .. controls (6.95,-1.4) and (3.31,-0.3) .. (0,0) .. controls (3.31,0.3) and (6.95,1.4) .. (10.93,3.29)   ;

\draw [color={rgb, 255:red, 126; green, 126; blue, 126 }  ,draw opacity=0.85 ]   (380.5,192.67) -- (453.63,270.22) ;
\draw [shift={(455,271.67)}, rotate = 226.68] [color={rgb, 255:red, 126; green, 126; blue, 126 }  ,draw opacity=0.85 ][line width=0.75]    (10.93,-3.29) .. controls (6.95,-1.4) and (3.31,-0.3) .. (0,0) .. controls (3.31,0.3) and (6.95,1.4) .. (10.93,3.29)   ;

\draw [color={rgb, 255:red, 126; green, 126; blue, 126 }  ,draw opacity=0.85 ]   (156.5,193) .. controls (143.7,221.57) and (142.53,240.43) .. (157.79,269.66) ;
\draw [shift={(158.5,271)}, rotate = 241.93] [color={rgb, 255:red, 126; green, 126; blue, 126 }  ,draw opacity=0.85 ][line width=0.75]    (10.93,-3.29) .. controls (6.95,-1.4) and (3.31,-0.3) .. (0,0) .. controls (3.31,0.3) and (6.95,1.4) .. (10.93,3.29)   ;

\draw [color={rgb, 255:red, 126; green, 126; blue, 126 }  ,draw opacity=0.85 ]   (262.5,193) .. controls (249.7,221.57) and (248.53,240.43) .. (263.79,269.66) ;
\draw [shift={(264.5,271)}, rotate = 241.93] [color={rgb, 255:red, 126; green, 126; blue, 126 }  ,draw opacity=0.85 ][line width=0.75]    (10.93,-3.29) .. controls (6.95,-1.4) and (3.31,-0.3) .. (0,0) .. controls (3.31,0.3) and (6.95,1.4) .. (10.93,3.29)   ;

\draw [color={rgb, 255:red, 126; green, 126; blue, 126 }  ,draw opacity=0.85 ]   (362.5,194) .. controls (349.7,222.57) and (348.53,242.4) .. (363.79,271.66) ;
\draw [shift={(364.5,273)}, rotate = 241.93] [color={rgb, 255:red, 126; green, 126; blue, 126 }  ,draw opacity=0.85 ][line width=0.75]    (10.93,-3.29) .. controls (6.95,-1.4) and (3.31,-0.3) .. (0,0) .. controls (3.31,0.3) and (6.95,1.4) .. (10.93,3.29)   ;

\draw [color={rgb, 255:red, 126; green, 126; blue, 126 }  ,draw opacity=0.85 ]   (467.5,193) .. controls (483.26,217.63) and (483.5,238.37) .. (470.12,269.57) ;
\draw [shift={(469.5,271)}, rotate = 293.63] [color={rgb, 255:red, 126; green, 126; blue, 126 }  ,draw opacity=0.85 ][line width=0.75]    (10.93,-3.29) .. controls (6.95,-1.4) and (3.31,-0.3) .. (0,0) .. controls (3.31,0.3) and (6.95,1.4) .. (10.93,3.29)   ;

\draw  [color={rgb, 255:red, 65; green, 117; blue, 5 }  ,draw opacity=1 ] (143,63.5) .. controls (143,62.67) and (143.67,62) .. (144.5,62) -- (398,62) .. controls (398.83,62) and (399.5,62.67) .. (399.5,63.5) -- (399.5,149.83) .. controls (399.5,150.66) and (398.83,151.33) .. (398,151.33) -- (144.5,151.33) .. controls (143.67,151.33) and (143,150.66) .. (143,149.83) -- cycle ;
\draw  [color={rgb, 255:red, 65; green, 117; blue, 5 }  ,draw opacity=1 ] (143,166.67) .. controls (143,166.3) and (143.3,166) .. (143.67,166) -- (398.83,166) .. controls (399.2,166) and (399.5,166.3) .. (399.5,166.67) -- (399.5,205.33) .. controls (399.5,205.7) and (399.2,206) .. (398.83,206) -- (143.67,206) .. controls (143.3,206) and (143,205.7) .. (143,205.33) -- cycle ;
\draw  [color={rgb, 255:red, 65; green, 117; blue, 5 }  ,draw opacity=1 ] (144.5,257.84) .. controls (144.5,257.56) and (144.73,257.33) .. (145,257.33) -- (489,257.33) .. controls (489.27,257.33) and (489.5,257.56) .. (489.5,257.84) -- (489.5,286.83) .. controls (489.5,287.11) and (489.27,287.33) .. (489,287.33) -- (145,287.33) .. controls (144.73,287.33) and (144.5,287.11) .. (144.5,286.83) -- cycle ;
\draw  [color={rgb, 255:red, 65; green, 117; blue, 5 }  ,draw opacity=1 ] (434.5,62.22) .. controls (434.5,61.73) and (434.9,61.33) .. (435.39,61.33) -- (486.61,61.33) .. controls (487.1,61.33) and (487.5,61.73) .. (487.5,62.22) -- (487.5,150.11) .. controls (487.5,150.6) and (487.1,151) .. (486.61,151) -- (435.39,151) .. controls (434.9,151) and (434.5,150.6) .. (434.5,150.11) -- cycle ;
\draw  [color={rgb, 255:red, 65; green, 117; blue, 5 }  ,draw opacity=1 ] (144,212.67) .. controls (144,212.3) and (144.3,212) .. (144.67,212) -- (399.83,212) .. controls (400.2,212) and (400.5,212.3) .. (400.5,212.67) -- (400.5,251.33) .. controls (400.5,251.7) and (400.2,252) .. (399.83,252) -- (144.67,252) .. controls (144.3,252) and (144,251.7) .. (144,251.33) -- cycle ;
\draw  [color={rgb, 255:red, 65; green, 117; blue, 5 }  ,draw opacity=1 ] (434,166.67) .. controls (434,166.3) and (434.3,166) .. (434.67,166) -- (487.83,166) .. controls (488.2,166) and (488.5,166.3) .. (488.5,166.67) -- (488.5,205.33) .. controls (488.5,205.7) and (488.2,206) .. (487.83,206) -- (434.67,206) .. controls (434.3,206) and (434,205.7) .. (434,205.33) -- cycle ;
\draw  [color={rgb, 255:red, 65; green, 117; blue, 5 }  ,draw opacity=1 ] (434,212.67) .. controls (434,212.3) and (434.3,212) .. (434.67,212) -- (487.83,212) .. controls (488.2,212) and (488.5,212.3) .. (488.5,212.67) -- (488.5,251.33) .. controls (488.5,251.7) and (488.2,252) .. (487.83,252) -- (434.67,252) .. controls (434.3,252) and (434,251.7) .. (434,251.33) -- cycle ;

\draw (214,122) node   {$\vdots $};
\draw (339,122) node   {$\vdots $};
\draw (420,122) node   {$\vdots $};
\draw (169,139) node [scale=0.7]  {$( 2,K,0)$};
\draw (267,139) node [scale=0.7]  {$( 2,K-1,0)$};
\draw (340,134) node   {$\dotsc $};
\draw (376,139) node [scale=0.7]  {$( 2,1,0)$};
\draw (458,139) node [scale=0.7]  {$( 2,0,0)$};
\draw (169,185) node [scale=0.7]  {$( 1,K,0)$};
\draw (267,185) node [scale=0.7]  {$( 1,K-1,0)$};
\draw (340,181) node   {$\dotsc $};
\draw (376,185) node [scale=0.7]  {$( 1,1,0)$};
\draw (457,184) node [scale=0.7]  {$( 1,0,0)$};
\draw (170,231) node [scale=0.7]  {$( 0,K,0)$};
\draw (267,231) node [scale=0.7]  {$( 0,K-1,0)$};
\draw (340,226) node   {$\dotsc $};
\draw (376,231) node [scale=0.7]  {$( 0,1,0)$};
\draw (457,231) node [scale=0.7]  {$( 0,0,0)$};
\draw (170,278) node [scale=0.7]  {$( 0,K,1)$};
\draw (267,278) node [scale=0.7]  {$( 0,K-1,1)$};
\draw (340,274) node   {$\dotsc $};
\draw (376,278) node [scale=0.7]  {$( 0,1,1)$};
\draw (457,278) node [scale=0.7]  {$( 0,0,1)$};
\draw (171,77) node [scale=0.7]  {$( K,K,0)$};
\draw (269,77) node [scale=0.7]  {$( K,K-1,0)$};
\draw (340,73) node   {$\dotsc $};
\draw (377,77) node [scale=0.7]  {$( K,1,0)$};
\draw (461,77) node [scale=0.7]  {$( K,0,0)$};
\draw (466,269) node [scale=0.5,color={rgb, 255:red, 208; green, 2; blue, 27 }  ,opacity=1 ]  {${\displaystyle 0}$};
\draw (383,270) node [scale=0.5,color={rgb, 255:red, 208; green, 2; blue, 27 }  ,opacity=1 ]  {$1$};
\draw (281,270) node [scale=0.5,color={rgb, 255:red, 208; green, 2; blue, 27 }  ,opacity=1 ]  {$K-1$};
\draw (180,270) node [scale=0.5,color={rgb, 255:red, 208; green, 2; blue, 27 }  ,opacity=1 ]  {$K$};
\draw (450,241) node [scale=0.5,color={rgb, 255:red, 208; green, 2; blue, 27 }  ,opacity=1 ]  {$K+1$};
\draw (365,223) node [scale=0.5,color={rgb, 255:red, 208; green, 2; blue, 27 }  ,opacity=1 ]  {$K+2$};
\draw (277,223) node [scale=0.5,color={rgb, 255:red, 208; green, 2; blue, 27 }  ,opacity=1 ]  {$2K$};
\draw (185,223) node [scale=0.5,color={rgb, 255:red, 208; green, 2; blue, 27 }  ,opacity=1 ]  {$2K+1$};
\draw (475,176) node [scale=0.5,color={rgb, 255:red, 208; green, 2; blue, 27 }  ,opacity=1 ]  {$2K+2$};
\draw (363,177) node [scale=0.5,color={rgb, 255:red, 208; green, 2; blue, 27 }  ,opacity=1 ]  {$2K+3$};
\draw (257,177) node [scale=0.5,color={rgb, 255:red, 208; green, 2; blue, 27 }  ,opacity=1 ]  {$3K+1$};
\draw (157,177) node [scale=0.5,color={rgb, 255:red, 208; green, 2; blue, 27 }  ,opacity=1 ]  {$3K+2$};
\draw (158,131) node [scale=0.5,color={rgb, 255:red, 208; green, 2; blue, 27 }  ,opacity=1 ]  {$4K+3$};
\draw (167,67) node [scale=0.5,color={rgb, 255:red, 208; green, 2; blue, 27 }  ,opacity=1 ]  {$( K+1)^{2} +K$};
\draw (253,131) node [scale=0.5,color={rgb, 255:red, 208; green, 2; blue, 27 }  ,opacity=1 ]  {$4K+2$};
\draw (364,131) node [scale=0.5,color={rgb, 255:red, 208; green, 2; blue, 27 }  ,opacity=1 ]  {$3K+4$};
\draw (450,131) node [scale=0.5,color={rgb, 255:red, 208; green, 2; blue, 27 }  ,opacity=1 ]  {$3K+3$};
\draw (268,67) node [scale=0.5,color={rgb, 255:red, 208; green, 2; blue, 27 }  ,opacity=1 ]  {$( K+1)^{2} +K-1$};
\draw (374,67) node [scale=0.5,color={rgb, 255:red, 208; green, 2; blue, 27 }  ,opacity=1 ]  {$( K+1)^{2} +1$};
\draw (452,68) node [scale=0.5,color={rgb, 255:red, 208; green, 2; blue, 27 }  ,opacity=1 ]  {$( K+1)^{2}$};
\draw (166,55) node [scale=0.7] [align=left] {Eq.~\eqref{8}};
\draw (135,185) node [scale=0.7,rotate=-270] [align=left] {Eq.~\eqref{9}};
\draw (136,231) node [scale=0.7,rotate=-270] [align=left] {Eq.~\eqref{10}};
\draw (461,54) node [scale=0.7] [align=left] {Eq.~\eqref{11}};
\draw (496,185) node [scale=0.7,rotate=-270] [align=left] {Eq.~\eqref{12}};
\draw (496,231) node [scale=0.7,rotate=-270] [align=left] {Eq.~\eqref{13}};
\draw (450,294) node [scale=0.7] [align=left] {$\text{Absorbing States}$};
\end{tikzpicture}
    \end{center}
\vspace{-5mm}
\caption{State transition diagram for $\mathcal{M}$ (self-transition loops have been omitted and only the states that can be reached with a non-zero probability are represented). Each state has been tagged with its numeric label.}
    \label{fig.amc}
    \vspace{-5mm}
\end{figure}

We derive the probability of Eve recovering the source message, i.e., the intercept probability, by means of the Markov chain $\mathcal{M}$ (shown in Fig.~\ref{fig.amc}) where its states are defined as follows.
\begin{definition}\label{def.state} 
We say that $\mathcal{M}$ is in state $(d_\mathrm{B},d_\mathrm{E},\delta)$ if $\mathrm{def}(\mathbf{M}_{\mathrm{B}}) = d_\mathrm{B}$, $\mathrm{def}(\mathbf{M}_{\mathrm{E}}) = d_\mathrm{E}$, and the ACK has not ($\delta = 0$) or has been ($\delta = 1$) successfully received by Alice.
\end{definition}

{From Definition~\ref{def.state}, we observe that the total number of states defining $\mathcal{M}$ is $2(K+1)^2$, which directly follows from the fact that: (i) the maximum value of defect $d_\mathrm{B}$ and $d_\mathrm{E}$ is equal to $K$ (corresponding to the cases when Bob and Alice have not successfully received any coded packet), and (ii) a ACK can either be received ($\delta = 1$) or not ($\delta = 0$).

After a coded packet transmission, assuming $d_{\mathrm{B}} \geq 1$, the rank of $\mathbf{M}_{\mathrm{B}}$ will increase by one if and only if Bob receives a coded packet that is linearly independent with the previously received. Equivalently, the rank of $\mathbf{M}_{\mathrm{B}}$ can at most be increased by one after a single coded packet transmission, i.e., the defect of $\mathbf{M}_{\mathrm{B}}$ can at most be reduced by one after a coded packet transmission.
The same holds true from Eve.
As for the value of $\delta$, Bob will attempt to acknowledge the successful recovery of a source message as soon as $d_{\mathrm{B}}$ becomes equal to $0$. For these reasons, all the $K(K+1)$ states where $d_\mathrm{B} \geq 1$ and $\delta = 1$ cannot be reached and can be disregarded. Thus, we will only consider the remaining $2(K+1)^2 - K(K+1) = (K+1)\cdot(K+2)$ states.
\begin{example}
Assume the system is in state $(K,K,0)$ and ignore self-transition loops, Fig.~\ref{fig.amc} shows that $\mathcal{M}$ is expected to exhibits non-null transition probabilities for states $(K-1,K,0)$, $(K-1,K-1,0)$ and $(K,K-1,0)$ corresponding to the cases when Bob, Bob and Alice or just Alice successfully receive a linearly independent coded packet, respectively. Since Bob cannot transmit an ACK message before a source message has been recovered, the transition probability toward any state where $\delta=1$ is zero.
\end{example}}

We then label the remaining $(K+1)\cdot(K+2)$ states.
\begin{definition}\label{def.labeling}
Each state takes a numeric label ranging from $0$ to $(K+1)\cdot(K+2) - 1$. If $\delta = 1$, the label of a state is equal to $d_\mathrm{E}$, otherwise it is equal to $(d_\mathrm{B} + 1)(K+1) + d_\mathrm{E}$.
\end{definition}
Furthermore, in order to derive the probability transition matrix of $\mathcal{M}$, we prove the following lemma.
\begin{lemma}\label{lem.trans}
Assume that matrix $\mathbf{M}_{\mathrm{X}}$ consists of $K \times (t+1)$ elements and assume that the first $t$ columns are linearly independent, for $X \in \{\mathrm{B},\mathrm{E}\}$ and $1 \leq t \leq (K-1)$. If $p > \frac{1}{q}$, the probability $\mathrm{W}_t$ of $\mathbf{M}_{\mathrm{X}}$ having rank $t+1$ can be approximated as follows:
\begin{equation}
\mathrm{W}_t \cong (1-p^K) \exp\left(-\sum_{\ell = 2}^{t+1}\binom{t}{\ell - 1} \frac{\pi_{\ell,K}}{(1-p^K)^\ell}\right), \label{eq.lm.1}
\end{equation}
where
$\pi_{1,r} = \rho_{1,r}$, $\pi_{\ell,r} = \rho_{c,r} - \sum_{s = 1}^{\ell - 1} \binom{\ell - 1}{s} \rho_{s,\ell}\pi_{\ell-s,r}$
and
$\rho_{c,r} = \left[\frac{1}{q} \left(1+(q-1)\left(1-\frac{q(1-p)}{q-1}\right)\right)^c\right]^r.$ If $p = \frac{1}{q}$, $\mathrm{W}_t$ is
$\mathrm{W}_t = 1-\frac{1}{q^{K-t}}$.
\end{lemma}
\begin{IEEEproof}
Let $\mathrm{R}_{K,t+1} = \mathbb{P}\left[\mathrm{rank}(\mathbf{M}_{\mathrm{X}})\right]$ be the probability of matrix $\mathbf{M}_{\mathrm{X}}$ having rank $t+1$. That is, let $\mathbf{M}_{\mathrm{X},t}$ be the $K \times t$ matrix defined by the first $t$ columns of $\mathbf{M}_{\mathrm{X}}$. The relation
\begin{equation}\label{eq.app.0}
\mathrm{W}_t = \frac{\mathbb{P}\left[\mathrm{rank}(\mathbf{M}_{\mathrm{X}}) = t+1\right]}{\mathbb{P}\left[\mathrm{rank}(\mathbf{M}_{\mathrm{X},t}) = t\right]}
\end{equation}
 holds true due to the fact that if $\mathbf{M}_{\mathrm{X}}$ has rank $t+1$ then the first $t$ columns are linearly independent.
From~\cite[Theorem~3.1]{8248799}, in the case of a $r \times c$ sparse random matrix over $\mathbb{F}_q$, it follows that 
\begin{equation}\label{eq.6}
\mathrm{R}_{r,c} \cong (1 - p^r)^c \exp\left(-\sum_{\ell = 2}^c\binom{c}{\ell}\frac{\pi_{\ell,r}}{(1-p^r)^\ell}\right),
\end{equation}
for $r \geq c$. Thus, by substituting~\eqref{eq.6} in~\eqref{eq.app.0} and by noting that {$\binom{t}{\ell-1} = \binom{t+1}{\ell} - \binom{t}{\ell}$},~\eqref{eq.lm.1} holds.
Finally, the case when $p = 1/q$ directly follows form~\cite[Eq.~(2)]{8248799}.
\end{IEEEproof}

From~\eqref{eq.lm.1}, the probability transition matrix $\mathbf{P}$ of $\mathcal{M}$ can be approximated by means of the following lemma.
\begin{lemma}\label{eq.lem.P}
The probability $\mathrm{P}_{i,j}$ of moving from state $i$ to state $j$ can be approximated as follows (only non-zero probabilities are listed):
\begin{itemize}
\item If $(K+1)(K-\tau+2) - K \leq i \leq (K+1)(K-\tau+2) - 1$, for $\tau = 0, \ldots, (K-2)$,
\setlength{\arraycolsep}{0.0em}
  \vspace{-2mm}\begin{equation}\label{8}{\footnotesize
  \hspace{-7mm}\mathrm{P}_{i,j}\!\! \cong\!\! \left\{
  \begin{array}{l l}
      \epsilon_\mathrm{B}(1\!-\!\epsilon_\mathrm{E})\mathrm{W}_{K-d_\mathrm{E}} & \hspace{0mm}\text{if $j = i -1 \wedge d_\mathrm{B} \geq d_\mathrm{E}$}\\
      (1\!-\!\epsilon_\mathrm{E})[\mathrm{W}_{K-d_\mathrm{E}}\!\!-\!\!(1\!-\!\epsilon_\mathrm{E}) \mathrm{W}_{K-d_\mathrm{B}}] & \hspace{0mm}\text{if $j = i -1 \wedge d_\mathrm{B} < d_\mathrm{E}$}\\
      \epsilon_\mathrm{E}(1\!-\!\epsilon_\mathrm{B})\mathrm{W}_{K-d_\mathrm{B}} & \hspace{0mm}\text{if $j = i - K-1 \wedge d_\mathrm{E} \geq d_\mathrm{B}$}\\
     (1\!-\!\epsilon_\mathrm{B})[\mathrm{W}_{K-d_\mathrm{B}}\!\! -\!\!(1\!-\!\epsilon_\mathrm{B})\mathrm{W}_{K-d_\mathrm{E}}] & \hspace{0mm}\text{if $j = i - K-1 \wedge d_\mathrm{E} < d_\mathrm{B}$}\\
     (1\!-\!\epsilon_\mathrm{B})(1\!-\!\epsilon_\mathrm{E})\mathrm{W}_{K-\min(d_\mathrm{B},d_\mathrm{E})} & \text{if $j = i - K - 2$}\\
     1 -  \sum_{\stackrel{j = \{i-1,i-K-1,}{i-K-2\}}} \mathrm{P}_{i,j} & \text{if $j = i$}\\
  \end{array} \right.}
\end{equation}

\vspace{-2mm}\item If $2K+3 \leq i \leq 3K+2$,
\setlength{\arraycolsep}{0.0em}
\vspace{-1mm}\begin{equation}\label{9}{\footnotesize
  \hspace{-7mm}\mathrm{P}_{i,j}\!\! \cong\!\! \left\{
  \begin{array}{l l}
      \epsilon_\mathrm{B}(1\!-\!\epsilon_\mathrm{E})\mathrm{W}_{K-d_\mathrm{E}} & \hspace{-4mm}\text{if $j = i -1\wedge\, d_\mathrm{B} \geq d_\mathrm{E}$}\\
      (1\!-\!\epsilon_\mathrm{E})[\mathrm{W}_{K-d_\mathrm{E}}\!\!-\!\!(1\!-\!\epsilon_\mathrm{E}) \mathrm{W}_{K-d_\mathrm{B}}] & \hspace{-4mm}\text{if $j = i -1\wedge\, d_\mathrm{B} < d_\mathrm{E}$}\\
      \epsilon_\mathrm{K}\epsilon_\mathrm{E}(1\!-\!\epsilon_\mathrm{B})\mathrm{W}_{K-d_\mathrm{B}} & \hspace{-10mm}\text{if $j = i - K-1\wedge\,d_\mathrm{E} \geq d_\mathrm{B}$}\\
     \epsilon_\mathrm{K}(1\!-\!\epsilon_\mathrm{B})[\mathrm{W}_{K-d_\mathrm{B}}\!\! -\!\!(1\!-\!\epsilon_\mathrm{B})\mathrm{W}_{K-d_\mathrm{E}}] & \hspace{1mm}\text{if $j = i - K-1$}\\
    &\hspace{3mm}\text{$\wedge\,d_\mathrm{E} < d_\mathrm{B}$}\\
     \epsilon_\mathrm{K}(1\!-\!\epsilon_\mathrm{B})(1\!-\!\epsilon_\mathrm{E})\mathrm{W}_{K-\min(d_\mathrm{B},d_\mathrm{E})} & \hspace{1mm}\text{if $j = i - K - 2$}\\
      (1\!-\!\epsilon_\mathrm{K})\epsilon_\mathrm{E}(1\!-\!\epsilon_\mathrm{B})\mathrm{W}_{K-d_\mathrm{B}} & \hspace{-10mm}\text{if $j = i - 2K-2\wedge\,d_\mathrm{E} \geq d_\mathrm{B}$}\\
     \!\!(1\!-\!\epsilon_\mathrm{K})(1\!-\!\epsilon_\mathrm{B})[\mathrm{W}_{K-d_\mathrm{B}}\!\! -\!\!(1\!-\!\epsilon_\mathrm{B})\mathrm{W}_{K-d_\mathrm{E}}] & \hspace{1mm}\text{if $j = i - 2K-2$}\\
    & \hspace{3mm}\text{$\wedge\,d_\mathrm{E} < d_\mathrm{B}$}\\
     (1\!-\!\epsilon_\mathrm{K})(1\!-\!\epsilon_\mathrm{B})(1\!-\!\epsilon_\mathrm{E})\mathrm{W}_{K-\min(d_\mathrm{B},d_\mathrm{E})} & \hspace{1mm}\text{if $j = i - 2K - 3$}\\        
     1 -  \sum_{\stackrel{j = \{i-1,i-K-1,i-K-2,}{i-2K-2,i-2K-3\}}} \mathrm{P}_{i,j} & \hspace{1mm}\text{if $j = i$}\\
  \end{array} \right.}
\end{equation}

\item If $K+2 \leq i \leq 2K+1$,
\setlength{\arraycolsep}{0.0em}
\vspace{-1mm}\begin{equation}\label{10}{\footnotesize
  \hspace{-7mm}\mathrm{P}_{i,j}\!\! \cong\!\! \left\{
  \begin{array}{l l}
      \epsilon_\mathrm{K}(1-\epsilon_\mathrm{E})\mathrm{W}_{K-d_\mathrm{E}} & \quad\text{if $j = i-1$}\\
      (1-\epsilon_\mathrm{K})(1-\epsilon_\mathrm{E})\mathrm{W}_{K-d_\mathrm{E}} & \quad\text{if $j = i-K-1$}\\
      (1-\epsilon_\mathrm{K})[1-(1-\epsilon_\mathrm{E})\mathrm{W}_{K-d_\mathrm{E}}] & \quad\text{if $j = i-K-2$}\\
    \epsilon_\mathrm{K}[1-(1-\epsilon_\mathrm{E})\mathrm{W}_{K-d_\mathrm{E}}] & \quad\text{if $j = i$}\\
  \end{array} \right.}
\end{equation}

\item If $i = (K+1)(K-\tau+1)$, for $\tau = 0, \ldots, (K-2)$,
\setlength{\arraycolsep}{0.0em}
\vspace{-1mm}\begin{equation}\label{11}{\footnotesize
  \hspace{-7mm}\mathrm{P}_{i,j} \!\!\cong\!\! \left\{
  \begin{array}{l l}
    (1-\epsilon_\mathrm{B})\mathrm{W}_{K-d_\mathrm{B}} & \quad\text{if $j = i - K - 1$}\\      
    1 - (1-\epsilon_\mathrm{B})\mathrm{W}_{K-d_\mathrm{B}} & \quad\text{if $j = i$}\\
  \end{array} \right.}
\end{equation}

\item If $i = 2(K+1)$,
\setlength{\arraycolsep}{0.0em}
\vspace{-1mm}\begin{equation}\label{12}{\footnotesize
  \hspace{-7mm}\mathrm{P}_{i,j} \!\!\cong\!\! \left\{
  \begin{array}{l l}
      (1-\epsilon_\mathrm{K})(1-\epsilon_\mathrm{B})\mathrm{W}_{K-d_\mathrm{B}} & \quad\text{if $j = i - 2K - 2$}\\      
    \epsilon_\mathrm{K}(1-\epsilon_\mathrm{B})\mathrm{W}_{K-d_\mathrm{B}} & \quad\text{if $j = i - K - 1$}\\      
    1-(1-\epsilon_\mathrm{B})\mathrm{W}_{K-d_\mathrm{B}} & \quad\text{if $j = i$}\\
  \end{array} \right.}
\end{equation}

\vspace{-1mm}\item For $i = K+1$,
\setlength{\arraycolsep}{0.0em}
\vspace{-2mm}\begin{equation}\label{13}{\footnotesize
  \hspace{-7mm}\mathrm{P}_{i,j} \!\!=\!\! \left\{
  \begin{array}{l l}
    (1-\epsilon_\mathrm{K}) & \quad\text{if $j = i-K-1$}\\
    \epsilon_\mathrm{K} & \quad\text{if $j = i$}
  \end{array} \right.}
\end{equation}

\vspace{-2mm}\item If $0 \leq i \leq K$, the state is and absorbing state and, hence, $\mathrm{P}_{i,j} = 1$. 
\end{itemize}
\end{lemma}
\begin{IEEEproof}
{We consider the case as per~\eqref{8}. In particular, we consider the case where $j = i-1$, which we can informally regard as the case where a state transition occurs \emph{horizontally}, from left to right (see Fig.~\ref{fig.amc}).}
As such, Bob will either not correctly receive a coded packet with probability $\epsilon_\mathrm{B}$ or he will receive a coded packet without reducing the defect of $\mathbf{M}_\mathrm{B}$. Conversely, Eve successfully receives a coded packet that reduces the defect of $\mathbf{M}_\mathrm{E}$. That is, 
\setlength{\arraycolsep}{0.0em}
\begin{eqnarray}
\mathrm{P}_{i,j} &{}={}& \epsilon_\mathrm{B}(1-\epsilon_\mathrm{E})\mathbb{P}[\mathrm{rank}(\mathrm{M}_\mathrm{E}) = K-d_\mathrm{E}]\\
 &&{}+{} (1-\epsilon_\mathrm{B})(1-\epsilon_\mathrm{E})\notag\\
 &&{}\cdot{}\mathbb{P}[\mathrm{rank}(\mathrm{M}_\mathrm{B}) = K-d_\mathrm{B}
 \wedge \,\mathrm{rank}(\mathrm{M}_\mathrm{E}) = K-d_\mathrm{E}+1],\notag
\end{eqnarray}
since $\mathbf{M}_\mathrm{B}$ and $\mathbf{M}_\mathrm{E}$ are statistically correlated. Thus, we have the following cases.
If $d_\mathrm{B} \geq d_\mathrm{E}$, the probability of $\mathrm{M}_\mathrm{B}$ not reducing its defect while $\mathrm{M}_\mathrm{E}$ does is expect to be small. Thus, the term 
$\mathbb{P}[\mathrm{rank}(\mathrm{M}_\mathrm{B}) = K-d_\mathrm{B}\wedge\mathrm{rank}(\mathrm{M}_\mathrm{E}) = K-d_\mathrm{E}+1]$ can be disregarded, and relation
$\mathrm{P}_{i,j} \geq \epsilon_\mathrm{B}(1-\epsilon_\mathrm{E})\mathbb{P}[\mathrm{rank}(\mathrm{M}_\mathrm{E}) = K-d_\mathrm{E}]$
holds.
If $d_\mathrm{B} < d_\mathrm{E}$, the term $\mathbb{P}[\mathrm{rank}(\mathrm{M}_\mathrm{B}) = K-d_\mathrm{B}\wedge\mathrm{rank}(\mathrm{M}_\mathrm{E}) = K-d_\mathrm{E}+1]$ can be approximated by subtracting the probability of $\mathbf{M}_\mathrm{B}$ reducing its defect from the probability of $d_\mathrm{E}$ being reduced as a result of a successfully received coded packet. From~\cite[Lemma~3.2]{8281108}, it follows that
$\mathrm{P}_{i,j} \geq \epsilon_\mathrm{B}(1-\epsilon_\mathrm{E})\mathbb{P}[\mathrm{rank}(\mathrm{M}_\mathrm{E}) = K-d_\mathrm{E}]
+\, (1-\epsilon_\mathrm{B})(1-\epsilon_\mathrm{E}) \Big(\mathbb{P}[\mathrm{rank}(\mathrm{M}_\mathrm{E}) = K-d_\mathrm{E}+1]
 -{}{} \,\mathbb{P}[\mathrm{rank}(\mathrm{M}_\mathrm{B}) = K-d_\mathrm{B}+1]\Big)$.
{The same reasoning holds true when $j = i - K - 1$ and we informally say that the transition occurs \emph{vertically}, from top to bottom. In that case, the third and fourth cases of~\eqref{8} follows by simply substituting $\mathrm{E}$ with $\mathrm{B}$ in the first and second cases of the same relation.
Let us now consider the situation where $j = i - K -2$, which corresponds to the case where both $\mathrm{M}_\mathrm{B}$ and $\mathrm{M}_\mathrm{E}$ reduce their defect as a result of a successfully received coded packet. In this case, we informally say that the transition  occurs \emph{diagonally}.} That is, both Bob and Eve successfully receive a coded packet with probability $(1-\epsilon_\mathrm{B})(1-\epsilon_\mathrm{E})$. Since $\mathbf{M}_\mathrm{B}$ and $\mathbf{M}_\mathrm{E}$ are statistically correlated, from~\cite[Lemma~3.2]{8281108}, it follows that $\mathrm{P}_{i,j}$ is upper-bounded by the product of $(1\!-\!\epsilon_\mathrm{B})(1\!-\!\epsilon_\mathrm{E})$ and the probability of $\mathrm{M}_\mathrm{t}$ reducing its defect, where the index $\mathrm{t} \in \{\mathrm{B},\mathrm{E}\}$ signifies the matrix with the smallest defect between $\mathrm{M}_\mathrm{B}$ and $\mathrm{M}_\mathrm{E}$. We then approximate $\mathrm{P}_{i,j}$ with the aforementioned upper-bound.

{As for the cases when $i$ fulfills the conditions for~\eqref{9}, from Fig.~\ref{fig.amc}, we observe that the probability of having a horizontal transition ($j = i - 1$) can be approximated as per the first and second case of~\eqref{8}. Once again, the probability of having a vertical transition can be approximated according to the third and fourth case of~\eqref{8} multiplied for $(1-\epsilon_\mathrm{K})$ or $\epsilon_\mathrm{K}$ if the transition leads to a state where the ACK message has ($\delta = 1$) or has not been successfully delivered ($\delta = 0$), respectively. The same reasoning holds true for the diagonal transitions.

When $i$ fulfil the conditions for~\eqref{10}, transition probability can be seen as a special case of~\eqref{8} where $\mathrm{W}_{K-d_\mathrm{B}}$ is $0$ as the defect of $\mathbf{M}_\mathrm{B}$ is $0$. Relations~\eqref{11} and~\eqref{12} are special cases of~\eqref{8} and~\eqref{9}, respectively, where only vertical transitions are considered and $\mathrm{W}_{K-d_\mathrm{E}}$ is $0$, as $d_\mathrm{E}$ is equal to $0$. When $i$ fulfills the condition for~\eqref{13}, both $d_\mathrm{B}$ and $d_\mathrm{E}$ are equal to $0$ -- thus, the system remains in the state $(K+1)$ for as long as the ACK message cannot be successfully delivered. Finally, the first $K + 1$ states are absorbing as Bob can successfully acknowledge to Alice the recovery of the source message and the transmission of coded packets is subsequently halted.}
\end{IEEEproof}

From Lemma~\ref{eq.lem.P}, it follows that $\mathcal{M}$ does not contain any cycles other than loops. For these reasons, $\mathbf{P}$ is a \mbox{$(K+2)\times(K+1)$} lower-triangular matrix with non-zero diagonal elements, which makes $\mathbf{P}$ invertible in the real field. Finally, The intercept probability can be obtained as follows.
\begin{theorem}\label{th.th}
For a given probability $p$ and a maximum number of coded packet transmissions $\Hat{N}$, the intercept probability $\mathrm{I}_{\Hat{N}}(p)$ can be approximated as
\begin{equation}
\mathrm{I}_{\Hat{N}}(p) \cong \sum_{\stackrel{j \in \small\{\tau(K+1),} {\text{for $\tau = 0, \ldots, (K+1)$}\small\}}} \!\!\!\!\!\!\!\!\!\!\!\!\!\!\mathbf{P}^{\Hat{N}}\left((K+1)^2 + K,j\right), \label{eq.th}
\end{equation}
where $\mathbf{P}^{\Hat{N}}\left(s,t\right)$ signifies the $(s,t)$-th element of the matrix $\mathbf{P}$ after it has been elevated to the power of $\Hat{N}$, for $s$ and $t = 0, \ldots, (K+1)^2 + K$.
\end{theorem}
\begin{IEEEproof}
The system starts with probability $1$ from the state with label $(K+1)^2 + K$, i.e., the system starts from state $(K,K,0)$ with probability $1$. The term $\mathrm{I}_{\Hat{N}}(p)$ is equal to the probability of the system being in any of the states having $d_\mathrm{E}$ equal to $0$, for a given $\Hat{N}$. From Definition~\ref{def.labeling}, we observe that states with labels $\tau(K+1)$, for $\tau = 0, \ldots, (K+1)$ are associated with  those cases where Eve successfully recovered the information message. That is,~\eqref{eq.th} holds.
\end{IEEEproof}

\vspace{-1.5mm}\section{Optimization Model}\label{sec.OM}
We define the Intercept Minimization (IM) problem as follows:
\vspace{-5mm}\begin{align}
      \text{IM} &  \quad  \min_{p} \,\,  \mathrm{I}_{\Hat{N}}(p) \label{IM.of}\\
    \text{s.t.} &  \quad \mathrm{D}_{\Hat{N}}(p) \geq \Hat{D} \label{IM.c1}
\end{align}
where $\mathrm{D}_{\Hat{N}}(p)$ signifies the probability of Bob recovering the source message. For a given value of $p$ and $\Hat{N}$, constraint~\eqref{IM.c1} ensures that Bob recovers the source message with at least probability $\Hat{D}$. {Form~\cite{7335581,8248799}, it follows that the average number of coded packet transmissions needed to recover a source messages increases as $p$ increases. Thus, not only Eve but also Bob is expected to require more coded packet transmissions to recover a source message. To prevent the IM problem to minimize the intercept probability by increasing the value of $p$ at the expense of the number of coded packet transmissions, constraint~\eqref{IM.c1} not only imposes a minimum threshold for the probability of Bob recovering a source message but also it ensures that a source message has to be recovered by $\Hat{N}$ coded packets transmissions. As such, if we consider the case where one coded packet transmission takes place in one-time slot, the proposed optimization framework ensures the delivery of a source message with a probability greater than or equal to $\Hat{D}$ in $\Hat{N}$ time slots or less.}

\begin{remark}
By following the same reasoning as in Theorem~\ref{th.th}, term $\mathrm{D}_{\Hat{N}}(p)$ can be approximated as $\sum_{j = 0}^{K+1} \mathbf{P}^{\Hat{N}}\left((K+1)^2 + K,j\right)$. However, as discussed in the proof of Lemma~\ref{eq.lem.P}, the proposed approximation of $\mathbf{P}_{i,j}$ is likely to over-estimate both $\mathrm{I}_{\Hat{N}}(p)$ and $\mathrm{D}_{\Hat{N}}(p)$ -- thus making approximation~\eqref{eq.th} an empirical upper-bound of the system intercept probability but leading to potentially overestimating the probability of Bob recovering the source message. For the sake of solving the IM problem, $\mathrm{D}_{\Hat{N}}(p)$ is approximated by directly employing~\eqref{eq.6}, as per~\cite[Eq.~(2), Theorem~3.1]{8248799}: 
\begin{equation}\label{eq.22}
\mathrm{D}_{\Hat{N}}(p) \cong \sum_{n = K}^{\Hat{N}} \binom{\Hat{N}}{n} (1-\epsilon_\mathrm{B})^n \epsilon^{\Hat{N}-n} \mathrm{R}_{n,K}.
\end{equation}
\end{remark}

The the IM problem can be solved as follows.
\begin{remark}
From~\eqref{eq.lm.1}, it follows that term $\sum_{\ell = 2}^{t+1}\binom{t}{\ell - 1} \frac{\pi_{\ell,K}}{(1-p^K)^\ell}$ is a non-decreasing function of $p$, which makes $\mathrm{W}_t$ a non-increasing function of $p$.
That is, for a given $\Hat{N}$, the higher $p$, the more unlikely it gets for the system to be in any of the states with label $\tau(K+1)$, for $\tau = 0, \ldots, (K+1)$, i.e., the more unlikely it gets for Eve to recover the source message.
In the following section, we will show how the proposed approximation for the intercept probability $\mathrm{I}_{\Hat{N}}(p)$ is largely a non-increasing function of $p$, for $q^{-1} \leq p < 1$ and $\epsilon_\mathrm{K} \geq 0.85$. Similarly,~\eqref{eq.6} is a non-increasing function of $p$, which makes~\eqref{eq.22} a non-increasing function as well.
For these reasons, the solution of the IM problem is given by the real root of $\mathrm{D}_{\Hat{N}}(p) - \Hat{D} = 0$, which can be derided by employing the bisection method.
\end{remark}

\vspace{-2mm}\section{Numerical Results}\label{sec.NR}
This section compares the derived expression of the intercept probability with Monte Carlo simulations, and solves the IM problem for different configurations. The code needed to reproduce our results is available online\footnote{\url{https://github.com/andreatassi/SparseRLNC}.}.

\begin{figure}[tb]
\vspace{-8mm}
\centering
\hspace*{-2mm}\subfloat[$q = 2$]{\label{fig.1.1}
    \includegraphics[width=0.51\columnwidth]{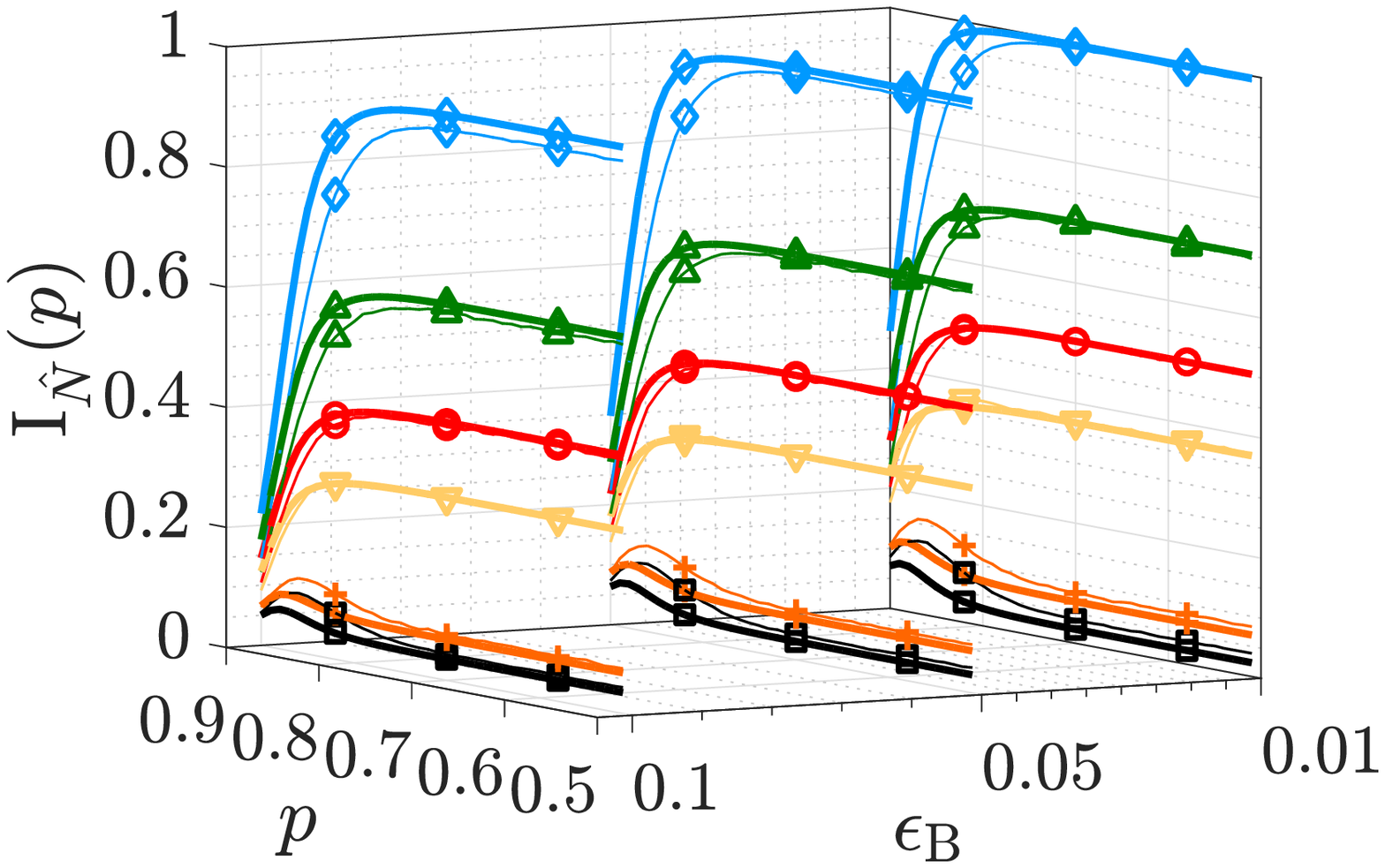}
}
\hspace*{-3mm}\subfloat[$q = 2^4$]{\label{fig.1.2}
    \includegraphics[width=0.51\columnwidth]{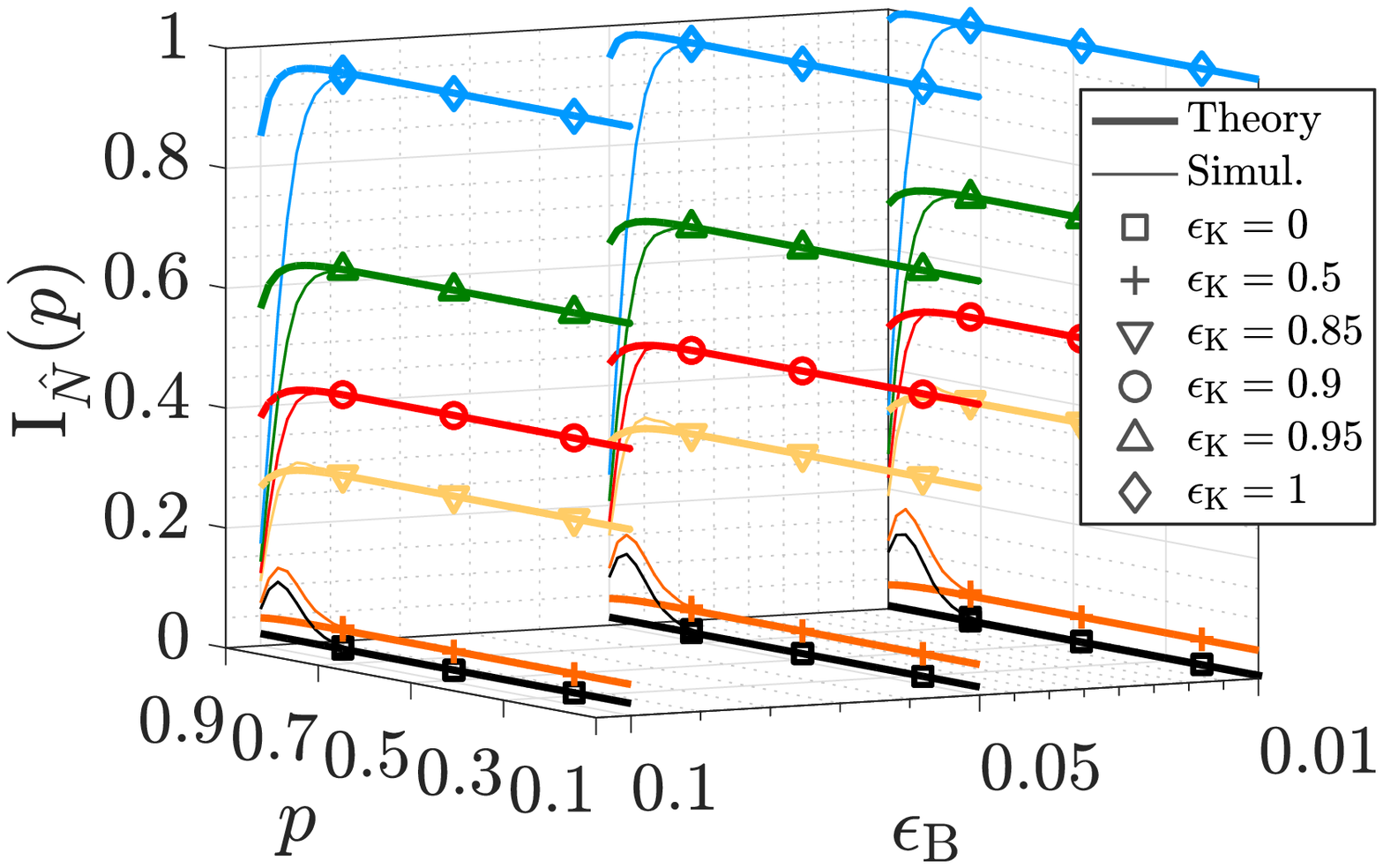}
}
\vspace{-1mm}
\caption{Comparison of $\mathrm{I}_\mathrm{\Hat{N}}(p)$ as a function of $p$ obtained through simulations (Simul.) and approximated (Theory) as in~\eqref{eq.th}, for $\epsilon_\mathrm{B} = \{0.01,0.05,0.1\}$, $\epsilon_\mathrm{E} = \epsilon_\mathrm{B} + 0.25$, $\epsilon_\mathrm{K} = \{0,0.5,0.85,0.9,0.95,1\}$, $K = 20$, $\Hat{N} = 2K$ and $q = \{2,2^4\}$. Legend of both figures is reported in Fig.~\ref{fig.1.2}.}\vspace{-5mm}
\label{fig.1}\vspace{-1mm}
\end{figure}

{Fig.~\ref{fig.1} compares the expression of the intercept probability as per~\eqref{eq.th} with Monte Carlo simulations, for $K = 20$, $q = \{2,2^4\}$ and $\Hat{N} = 2K$. We also set Bob's and Eve's packet error probability equal to $\epsilon_\mathrm{B} = \{0.01,0.05,0.1\}$ and $\epsilon_\mathrm{E} = \epsilon_\mathrm{B} + 0.25$, respectively.
In particular, Fig.~\ref{fig.1.1} shows that, for $q = 2$,~\eqref{eq.th} is a tight empirical approximation of the intercept probability -- the maximum Mean Squared Error (MSE) between simulations and our proposed approximation~\eqref{eq.th} is equal to $0.933\cdot 10^{-3}$, for $\epsilon_\mathrm{B} = 0.01$, $\epsilon_\mathrm{E} = 0.26$ and $\epsilon_\mathrm{K} = 1$.

For $q = 2^4$, Fig.~\ref{fig.1.2} shows that the intercept probability are almost constant for $2^{-4}\leq p \leq 0.73$, which follows from the fact that both $\rho_{c,r}$ and $\pi_{\ell,K}$ approach $0$ as $q$ grows (see Lemma~\ref{lem.trans}), and hence, $\mathrm{W}_t$ can be approximated with $(1-p^K)$. The proposed approximation becomes looser only when the probability $p$ of a source packet not taking part in the generation of coding vector is very large ($p \geq 0.8$).}

{From Fig.~\ref{fig.1}, we also observe that the proposed~\eqref{eq.th} is also an empirical upper-bound of the intercept probability both in the case of $q = 2$ and $2^4$, for $\epsilon_\mathrm{K} \geq 0.85$ and $\epsilon_\mathrm{K} \geq 0.9$, respectively.
In addition, for $p \geq 0.8$ and $\epsilon_\mathrm{K} \geq 0.85$, the simulated $\mathrm{I}_\mathrm{\Hat{N}}(p)$ sharply decreases as the value of $p$ approaches $0.9$ and hence, the probability of having all-zero coding vectors sharply increases thus, making for both Eve and Bob more unlikely to recover a source message -- for instance, if the value of $p$ increases from $0.8$ to $0.9$, the probability of having an all-zero coded packet increases from $0.012$ to $0.12$, for $K = 20$.
For $\epsilon_\mathrm{K} \leq 0.5$ or $\epsilon_\mathrm{K} \leq 0.85$, for $q = 2$ and $2^4$, respectively, the intercept probability increases with $p$, for $0.75 \leq p \leq 0.85$. That is, as $\epsilon_\mathrm{K}$ decreases, the number of coded packets transmitted after Bob has already recovered the source message decreases as well. This impacts on the probability of Eve recovering the source message, and hence, the overall value of $\mathrm{I}_\mathrm{\Hat{N}}(p)$ reduces up to $0.05$. In these cases, from Lemma~\ref{lem.trans}, we note that some composite transition probabilities are non-decreasing functions with $p$ and in this case they can be appreciated in the overall expression of $\mathrm{I}_\mathrm{\Hat{N}}(p)$. Assuming $\epsilon_\mathrm{E} = \epsilon_\mathrm{B} = 0$, $K = 20$ and that $\mathcal{M}$ transitions from $(4,5,0)$ to $(3,4,0)$ and then to $(3,3,0)$, the overall probability of this transitions to happen is $\mathrm{W}_{K-4}(\mathrm{W}_{K-4} - \mathrm{W}_{K-3})$ which is a non-decreasing function of $p$ when $0.7 \leq p \leq 0.87$ and $0.7 \leq p \leq 0.9$, for $q = 2$ and $2^4$, respectively.}

\begin{figure}[t!]
\vspace{-8mm}
\centering
\hspace*{-1.5mm}\subfloat[$\epsilon_\mathrm{B} = 0.05$, $\epsilon_\mathrm{E} = 0.2$]{\label{fig.2.1}
    \includegraphics[width=0.5\columnwidth]{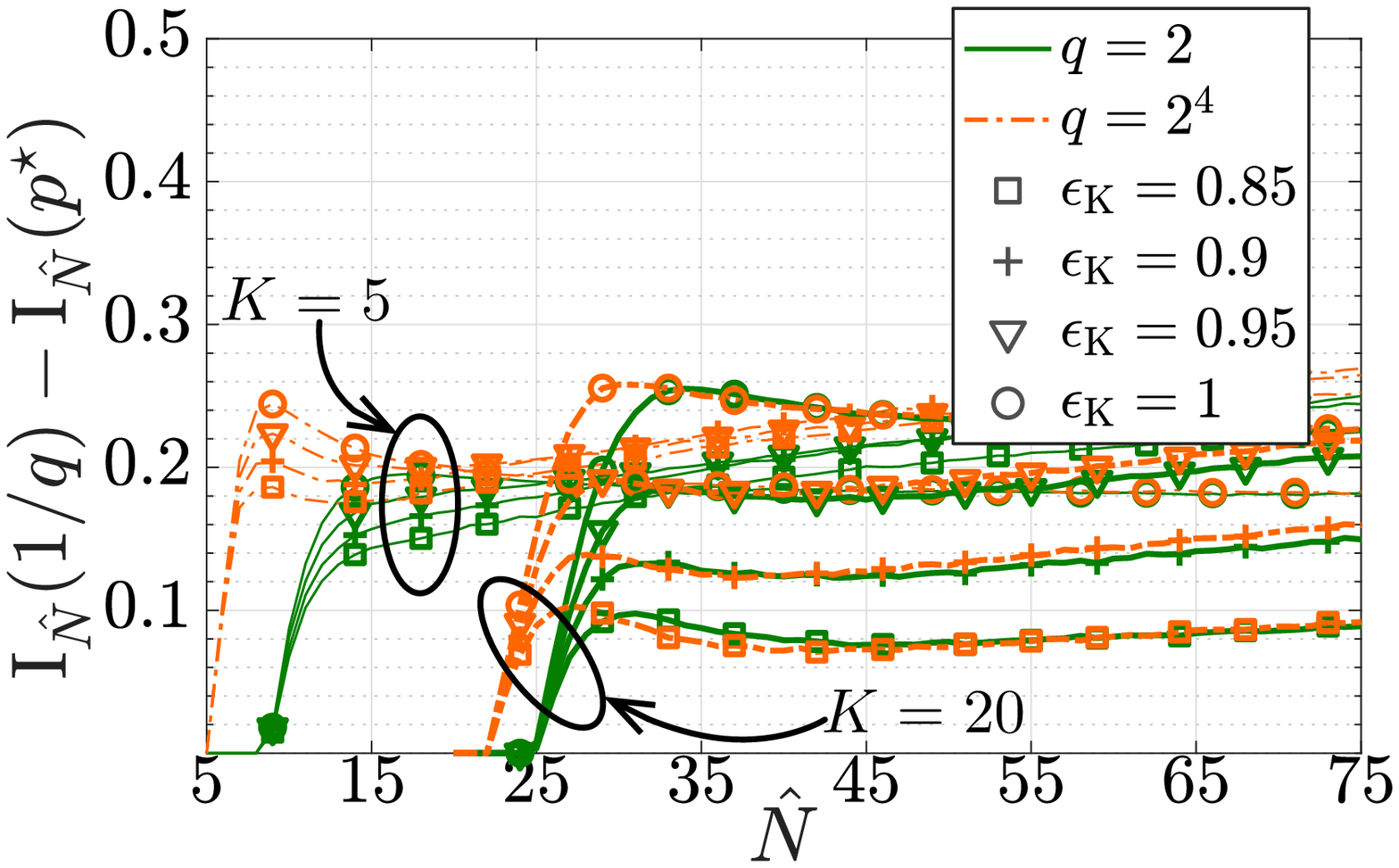}
}
\hspace*{-1.8mm}\subfloat[$\epsilon_\mathrm{B} = 0.05$, $\epsilon_\mathrm{E} = 0.3$]{\label{fig.2.2}
    \includegraphics[width=0.5\columnwidth]{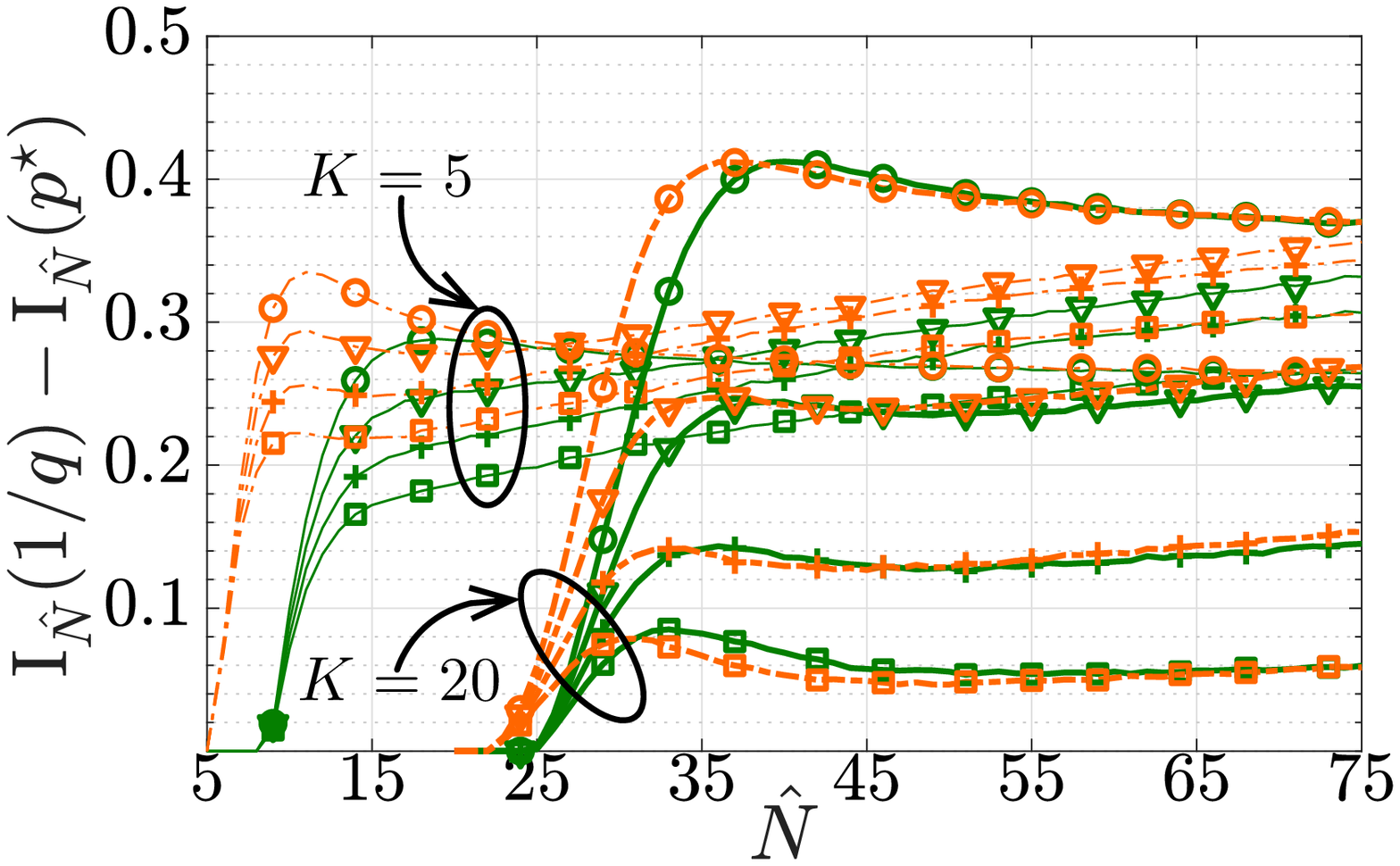}
}\\
\vspace*{-3mm}
\hspace*{-1.5mm}\subfloat[$\epsilon_\mathrm{B} = 0.1$, $\epsilon_\mathrm{E} = 0.25$]{\label{fig.2.3}
    \includegraphics[width=0.5\columnwidth]{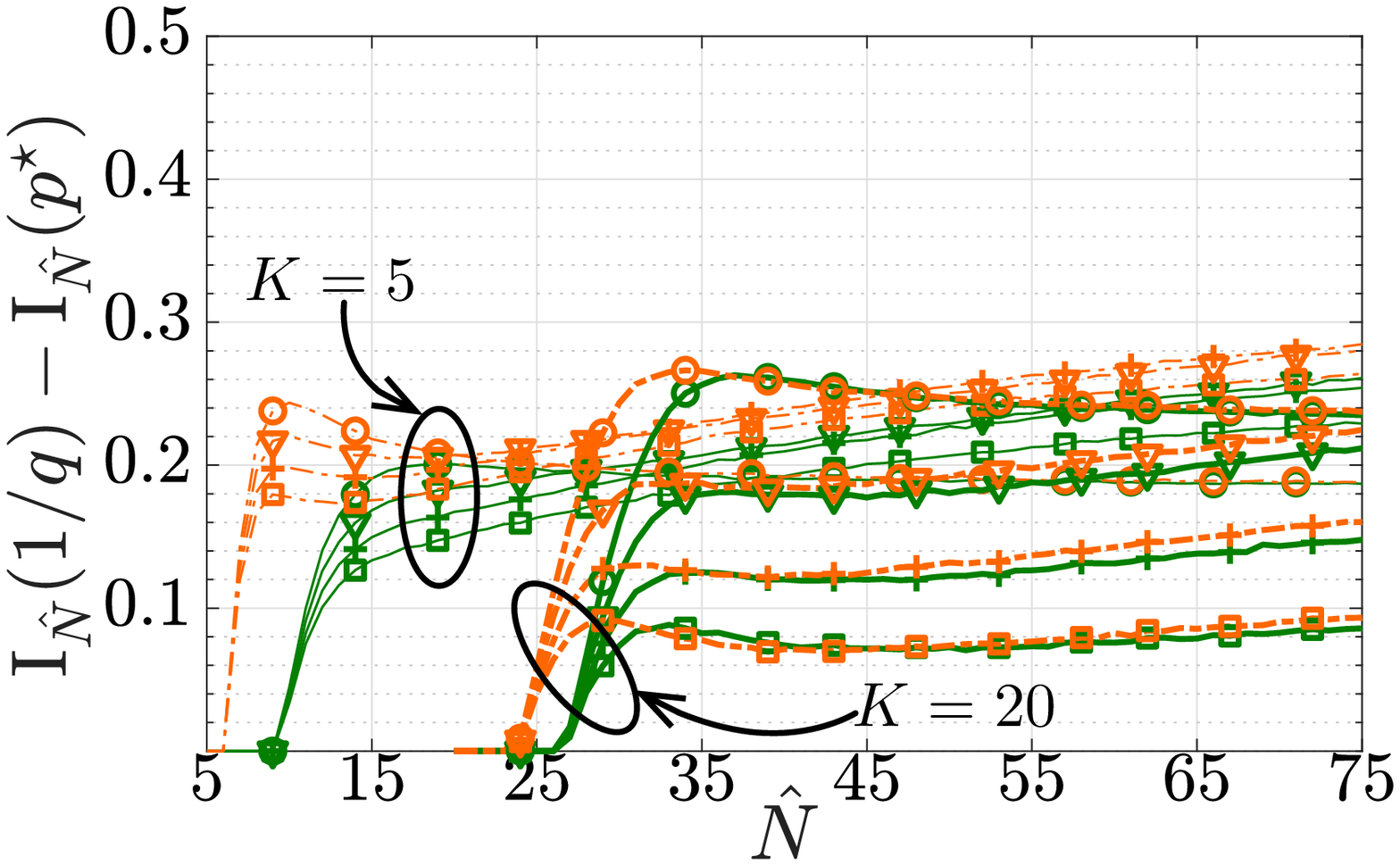}
}
\hspace*{-1.8mm}\subfloat[$\epsilon_\mathrm{B} = 0.1$, $\epsilon_\mathrm{E} = 0.35$]{\label{fig.2.4}
    \includegraphics[width=0.5\columnwidth]{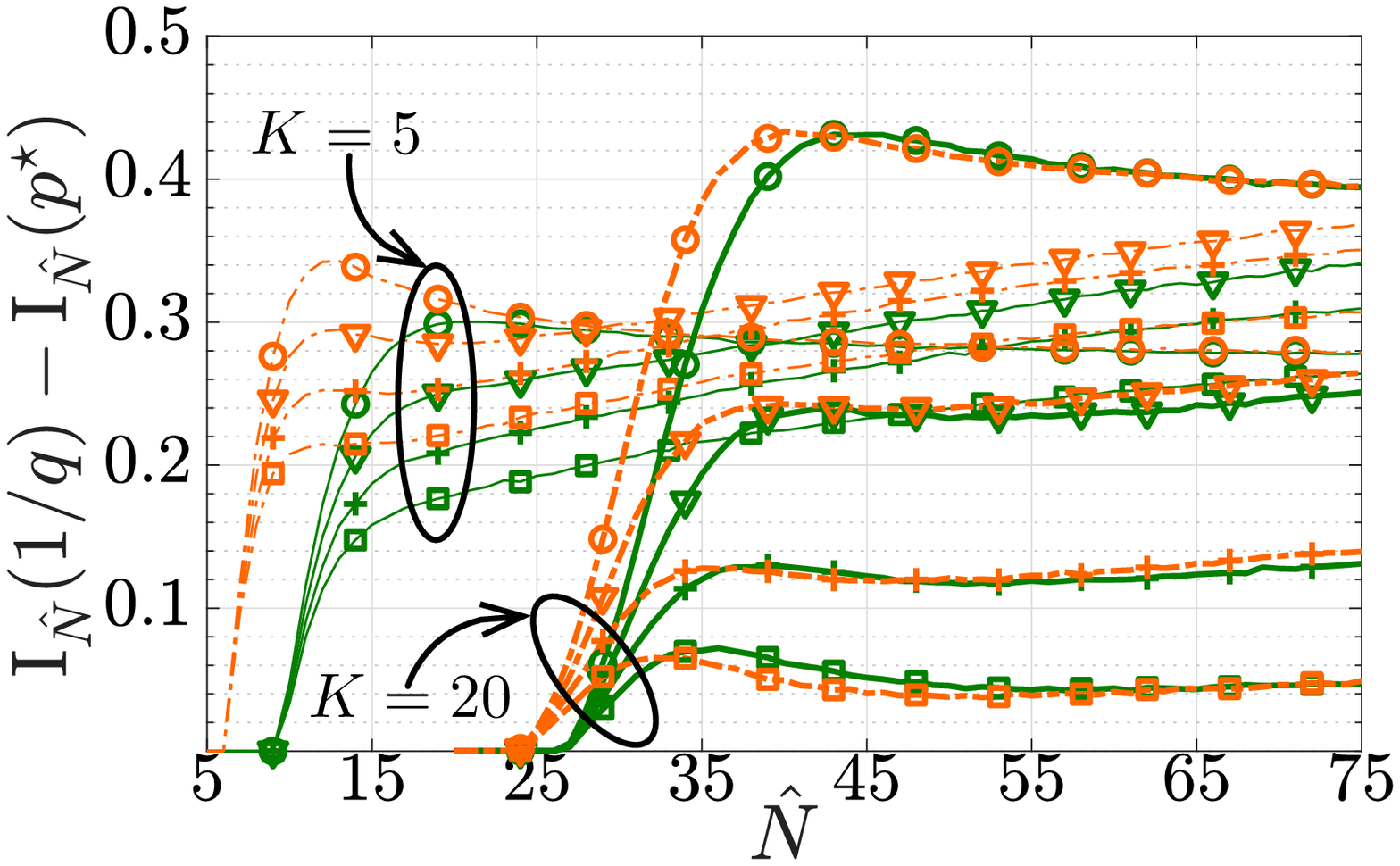}
}

\vspace{-0.4mm}
\caption{Intercept probability gain as a function of $\Hat{N}$, for $\epsilon_\mathrm{B} = \{0.05,0.1\}$, \mbox{$\epsilon_\mathrm{E} - \epsilon_\mathrm{B} = \{0.15, 0.25\}$}, $\epsilon_\mathrm{K} = \{0.85,0.9,0.95,1\}$, $K = \{5,20\}$ and $q = \{2,2^4\}$. Legend of all figures is reported in Fig.~\ref{fig.2.1}.}\vspace{-4mm}
\label{fig.2}
\end{figure}

{Fig.~\ref{fig.2} compares the intercept probability obtained by employing the proposed IM problem with the state-of-the-art performance of a system model as per~\cite{6777406,7214217} where $p = 1/q$ and hence, the classic RLNC is used. In particular, Fig.~\ref{fig.2} shows the \emph{intercept probability gain} defined as the difference between the intercept probability values obtained by using the classic RLNC and the intercept probability that we get by setting $p$ equal to the solution of the IM problem $p^\star$ -- namely, $\mathrm{I}_\mathrm{\Hat{N}}(1/q) - \mathrm{I}_\mathrm{\Hat{N}}(p^\star)$.}
In order to show the intercept probability gain effectively achieved, both $\mathrm{I}_\mathrm{\Hat{N}}(1/q)$ and $\mathrm{I}_\mathrm{\Hat{N}}(p^\star)$ are obtained by employing Monte Carlo simulations.

{Let us consider Fig.~\ref{fig.2.1}, for $\epsilon_\mathrm{B} = 0.05$, $\epsilon_\mathrm{E} = 0.2$, $K = 5$ and $q = 2$. In the case of $\epsilon_\mathrm{K} = 1$, the intercept probability gain sharply increases and reaches its maximum of $0.196$ for $\Hat{N} = 17$. As $\epsilon_\mathrm{K}$ decreases, the intercept probability gain decreases as well. In particular, for $\Hat{N} = 17$ and $\epsilon_\mathrm{K} = 0.85$, the intercept probability gain reduces to $0.15$. For $q = 2^4$, the intercept probability gain is generally larger. That is, for $\epsilon_\mathrm{K} = 1$ and $\epsilon_\mathrm{K} = 0.85$, the intercept probability gain reaches its maximum of $0.25$ and $0.27$, for $\Hat{N}=75$. With regard to Fig.~\ref{fig.2.2}, as $\epsilon_\mathrm{E}$ increases to $0.3$, the intercept probability gain reaches the value of $0.33$ and $0.36$, for $q = 2$ and $q = 2^4$, respectively. As $K$ is set equal to $20$, the intercept probability gain associated to $q = 2$ and $q = 2^4$ are comparable.} We also note that, as $\epsilon_\mathrm{K}$ decreases, we expect the intercept probability gain to decrease the chances of Eve successfully receiving enough coded packets to recover the source message are impaired by the reduced probability of Alice having to unnecessarily broadcast coded packets due to the loss of acknowledge messages from Bob.

{In Figs.~\ref{fig.2.3} and~\ref{fig.2.4}, Bob's packet error probability is doubled ($\epsilon_\mathrm{B} = 0.1$). Yet, the intercept probability gains are comparable to those in the cases where $\epsilon_\mathrm{B}$ was equal to $0.05$. Since in Fig.~\ref{fig.2} the difference $\epsilon_\mathrm{E} - \epsilon_\mathrm{B}$ is fixed and set equal to $0.15$ or $0.25$, we can conclude that the value of the intercept probability gain is determined by the difference in the packet error probability between Eve and Bob, for a given $p$ and $\Hat{N}$.}

\vspace{-3mm}\section{Conclusions}\label{sec.CL}
We present a novel strategy for approximating the intercept probability for networks where secrecy is achieved by employing a sparse implementation of RLNC. The proposed approximation is general and applies to the cases where transmissions are not acknowledged or when they are and the eavesdropper jams the feedback channel. We also propose an optimization framework for minimizing the intercept probability by increasing the sparsity of RLNC in use. Analytic results empirically establish that the proposed approximation for the intercept probability is tight, for practical network and transmission parameters. Our optimization framework ensures a reduction of the intercept probability of up to $82\%$ compared to the case where classic RLNC is used.

\vspace{-3mm}
\section{Acknowledgments}
The authors would like to thank Oliver Johnson (University of Bristol, Bristol, UK) for the insightful discussions and precious feedback.

\vspace{-6mm}
\bibliographystyle{IEEEtran}
\bibliography{IEEEabrv,papers}

\begin{thebibliography}{1}
\providecommand{\url}[1]{#1}
\csname url@samestyle\endcsname
\providecommand{\newblock}{\relax}
\providecommand{\bibinfo}[2]{#2}
\providecommand{\BIBentrySTDinterwordspacing}{\spaceskip=0pt\relax}
\providecommand{\BIBentryALTinterwordstretchfactor}{4}
\providecommand{\BIBentryALTinterwordspacing}{\spaceskip=\fontdimen2\font plus
\BIBentryALTinterwordstretchfactor\fontdimen3\font minus
  \fontdimen4\font\relax}
\providecommand{\BIBforeignlanguage}[2]{{%
\expandafter\ifx\csname l@#1\endcsname\relax
\typeout{** WARNING: IEEEtran.bst: No hyphenation pattern has been}%
\typeout{** loaded for the language `#1'. Using the pattern for}%
\typeout{** the default language instead.}%
\else
\language=\csname l@#1\endcsname
\fi
#2}}
\providecommand{\BIBdecl}{\relax}
\BIBdecl

\bibitem{4529264}
M.~Bloch, J.~Barros, M.~R.~D. Rodrigues, and S.~W. McLaughlin, ``{Wireless
  Information-Theoretic Security},'' \emph{{IEEE} Trans. Inf. Theory}, vol.~54,
  no.~6, pp. 2515--2534, Jun. 2008.

\bibitem{6777406}
H.~Niu, M.~Iwai, K.~Sezaki, L.~Sun, and Q.~Du, ``{Exploiting Fountain Codes for
  Secure Wireless Delivery},'' \emph{{IEEE} Commun. Lett.}, vol.~18, no.~5, May
  2014.

\bibitem{7214217}
A.~S. Khan, A.~Tassi, and I.~Chatzigeorgiou, ``{Rethinking the Intercept
  Probability of Random Linear Network Coding},'' \emph{{IEEE} Commun. Lett.},
  vol.~19, no.~10, Oct. 2015.

\bibitem{1023595}
N.~Cai and R.~W. Yeung, ``{Secure Network Coding},'' in \emph{Proc. of IEEE
  ISIT}, Lausanne, CH, Jun. 2002.

\bibitem{8281108}
E.~Tsimbalo, A.~Tassi, and R.~J. Piechocki, ``{Reliability of Multicast Under
  Random Linear Network Coding},'' \emph{{IEEE} Trans. Commun.}, vol.~66,
  no.~6, Jun. 2018.

\bibitem{7335581}
A.~Tassi, I.~Chatzigeorgiou, and D.~E. Lucani, ``{Analysis and Optimization of
  Sparse Random Linear Network Coding for Reliable Multicast Services},''
  \emph{{IEEE} Trans. Commun.}, vol.~64, no.~1, pp. 285--299, Jan. 2016.

\bibitem{bloch_barros_2011}
M.~Bloch and J.~Barros, \emph{{Physical-Layer Security: From Information Theory
  to Security Engineering}}.\hskip 1em plus 0.5em minus 0.4em\relax Cambridge
  University Press, 2011.

\bibitem{1055917}
S.~Leung-Yan-Cheong and M.~Hellman, ``{The Gaussian Wire-Tap Channel},''
  \emph{{IEEE} Trans. Inf. Theory}, vol.~24, no.~4, pp. 451--456, July 1978.

\bibitem{8248799}
S.~Brown, O.~Johnson, and A.~Tassi, ``{Reliability of Broadcast Communications
  Under Sparse Random Linear Network Coding},'' \emph{{IEEE} Trans. Veh.
  Technol.}, vol.~67, no.~5, May 2018.

\end{thebibliography}

\end{document}